# A Wideband STAR-RIS Employing Shared-Aperture Polarization Conversion and Diagonal Switching Topology for 6G Applications

Ruiqi Wang, *Member, IEEE* and Atif Shamim, *Fellow, IEEE*

***Abstract*—Simultaneously transmitting and reflecting reconfigurable intelligent surfaces (STAR-RIS) enable programmable wave control on both sides of a surface, extending wireless coverage beyond single-sided operation. However, practical STAR-RIS demonstrations in literature have narrow bandwidths, because it is challenging to achieve wideband reflection and transmission simultaneously from the same aperture. To overcome this limitation, a wideband STAR-RIS architecture is proposed based on shared-aperture polarization conversion and diagonal switching topology. Here, multiple resonances enable wideband magnitude responses and mirror-symmetric current reversal maintains 180° phase difference in both reflection and transmission modes. Accordingly, a diagonal four-branch unit cell configuration is developed, where the tailored top patch directs the received wave either through a pair of blind vias connected by a buried strip to the shared top aperture for reflection, or to the bottom ring patch using a through via for transmission. Based on the proposed design, a 10 × 10 STAR-RIS prototype is fabricated and experimentally characterized. The measured 3-dB bandwidths reach 19.8% in reflection mode and 21.9% in transmission mode. Near-field measurements show average signal enhancements of 19.8 and 22.6 dB for reflection and transmission, respectively, over the wide frequency range of 6.8–8.4 GHz. Far-field measurements further demonstrate average received-power enhancements of 11.2 dB in reflection and 11.9 dB in transmission. To the best of the authors' knowledge, this work presents the first experimentally demonstrated wideband STAR-RIS covering the 7.125–8.4 GHz band within the emerging 6G FR3 spectrum.**

## I. Introduction

Simultaneously transmitting and reflecting reconfigurable intelligent surfaces (STAR-RIS) have emerged as an extension of conventional reconfigurable intelligent surface (RIS) to enable programmable electromagnetic (EM) wave manipulation on both sides of a surface [1], [2]. Unlike reflection-only RIS, whose controllable coverage is restricted to the same side as the incident wave [3] – [16], STAR-RIS can manipulate both reflected and transmitted waves, thereby extending the service region from half-space to full-space [17], [18]. As illustrated in Fig. 1, when a STAR-RIS is deployed on a building window, the reflected wave can be steered toward outdoor users located in blocked regions, whereas the transmitted wave can provide coverage for indoor users. This capability is attractive to future 6G networks, where users may be distributed across different propagation regions and a single programmable surface is expected to provide flexible coverage on both sides of the deployment interface [19] – [22]. Among the spectrum ranges being explored for future 6G systems, the upper-mid-band spectrum, commonly referred to as frequency range 3 (FR3), spanning approximately 7.125–24.25 GHz, is of particular interest because it offers a favorable tradeoff between the propagation characteristics of lower-frequency bands and the larger available bandwidth at higher frequencies [23]. However, realizing practical STAR-RIS hardware in FR3 requires broadband reflection and transmission from the same aperture while maintaining high magnitude and stable phase quantization over the common operating band.

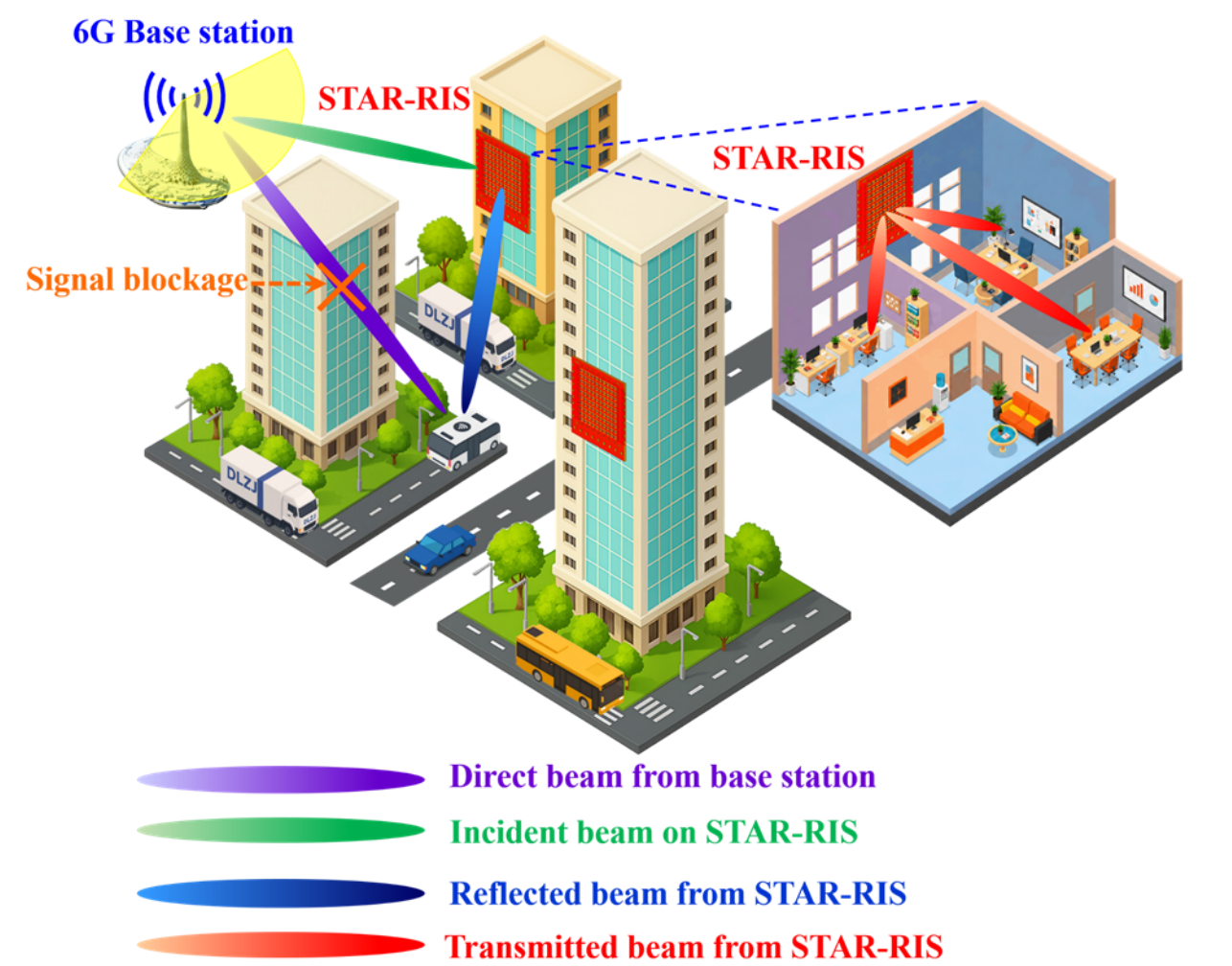

Fig. 1. Conceptual application scenario of the proposed STAR-RIS for outdoor reflection and indoor transmission.

The authors are with the Computer, Electrical and Mathematical Sciences and Engineering Division, King Abdullah University of Science and Technology (KAUST), Thuwal 23955, Saudi Arabia (e-mail: ruiqi.wang.1@kaust.edu.sa; atif.shamim@kaust.edu.sa)

In the literature, although STAR-RIS has been extensively investigated in the communication community, in terms of channel modeling, resource allocation, and transmission/reflection coefficient optimization [24], [25], experimental hardware demonstrations remain relatively limited, particularly for wideband operation. Several representative implementations have recently been reported. A 16 × 16 STAR reconfigurable array is designed in [26]; however, the measured gain bandwidth is 8.4%. In [27], an active STAR-RIS was experimentally demonstrated over 2.58–2.63 GHz, corresponding to an operating bandwidth of 1.9%. More recently, a hybrid transmitting and reflecting beyond-diagonal RIS is implemented at 2.4 GHz, where the 2-bit phase-reconfigurable antenna provides a bandwidth of 200 MHz, corresponding to 8.3% [28]. Moreover, the design in [28] relies on two phase-reconfigurable antenna arrays interconnected through tunable two-port power splitters, resulting in a multi-module architecture that increases the complexity of compact surface integration. These works demonstrate the feasibility of practical STAR-RIS hardware; however, their operating bandwidths remain relatively limited. Moreover, the main challenge is achieving wideband reflection and transmission simultaneously through a common aperture over the same frequency range. In addition, the RF routing and DC biasing should be integrated with minimal impact on the external EM response. Therefore, wideband, compact, and integrated STAR-RIS hardware targeting the emerging 6G FR3 spectrum remains insufficiently explored.

To bridge the research gap, this work develops a wideband diagonal-switching STAR-RIS targeting the 7.125–8.4 GHz band for future 6G FR3 applications [29] – [32]. The proposed four-state polarization-converting unit cell combines reconfigurable diagonal switching with a shared-aperture architecture to provide 1-bit phase control in both reflection and transmission modes. Wideband operation is achieved through multiple resonances for the magnitude response and mirror-symmetric current reversal for maintaining a stable phase difference. In addition, the RF routing and DC biasing networks are embedded within the multilayer structure to facilitate compact and scalable implementation. A 10 × 10 prototype is fabricated and experimentally validated, achieving measured 3-dB bandwidths of 19.8% and 21.9% for reflection and transmission, respectively. The remainder of this paper is organized as follows. Section II presents the unit-cell design and operating characteristics. Section III describes the array-level beamforming and control architecture. Section IV presents the fabricated prototype and experimental validation, followed by the conclusion in Section V.

## II. Unit Cell Design and Simulation

The configuration of the proposed STAR-RIS unit cell is shown in Fig. 2. The unit cell employs a diagonal topology with four switching branches to realize wideband reflection and transmission within a shared aperture, while providing two reconfigurable phase states for each operating mode. Four reconfigurable states are therefore obtained, including two states for 1-bit reflection phase quantization and two states for 1-bit transmission phase quantization. The unit cell adopts a multilayer PCB configuration that integrates the reflection aperture, transmission aperture, RF routing network, and DC biasing lines within a compact structure.

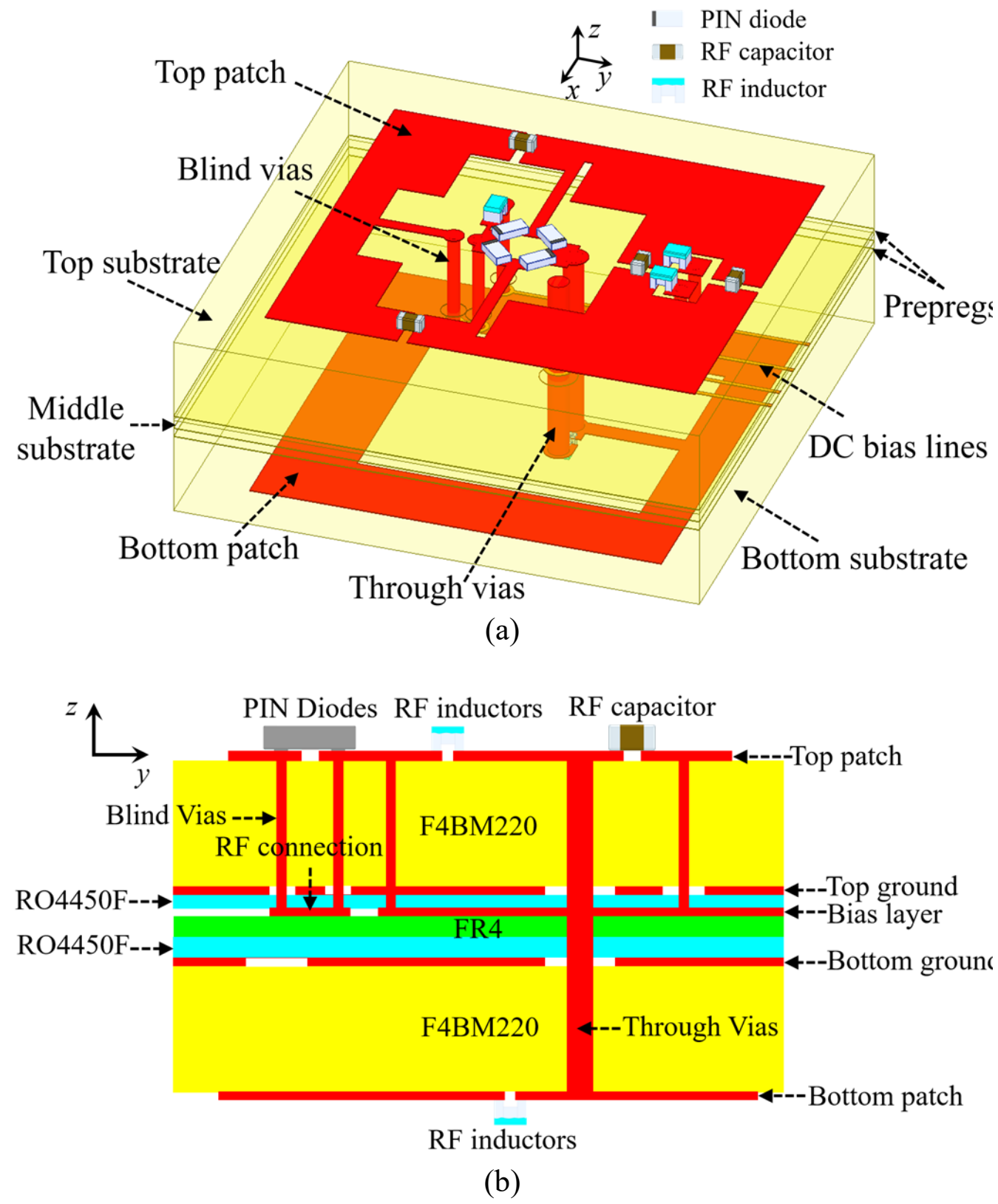

Fig. 2. Configuration of the proposed STAR-RIS unit cell. (a) 3-D view. (b) Cross-sectional view.

As illustrated in Fig. 2, the top radiating patch is printed on a 2-mm-thick F4BM220 substrate with a relative permittivity of 2.2 and a loss tangent of 0.001. The relatively thick, low-permittivity substrate is selected to enhance the operating bandwidth of the radiating structure. The two blind vias associated with the reflection path have a radius of 0.15 mm and provide RF connections between the top radiating structure and the buried routing layer through the top ground plane. Beneath the top ground plane, a 0.2-mm-thick FR4 layer with a relative permittivity of 4.4 and a loss tangent of 0.02 is introduced to accommodate the internal RF routing network. This layer is bonded to the F4BM220 substrates using 0.1-mm-thick RO4450F prepregs with a relative permittivity of 3.52 and a loss tangent of 0.0041. On the transmission side, a second 2-mm-thick F4BM220 substrate is employed, with the bottom ring patch printed on its outer surface to form the transmitting aperture. A through via with a radius of 0.25 mm extends through the complete multilayer structure and provides the RF connection to the bottom transmitting aperture.

A key feature of the proposed multilayer topology is that the internal RF routing network is fully enclosed between the top and bottom ground planes. This embedded configuration electromagnetically shields the routing network from the external radiating apertures, thereby reducing undesired interaction with both the reflection and transmission responses.

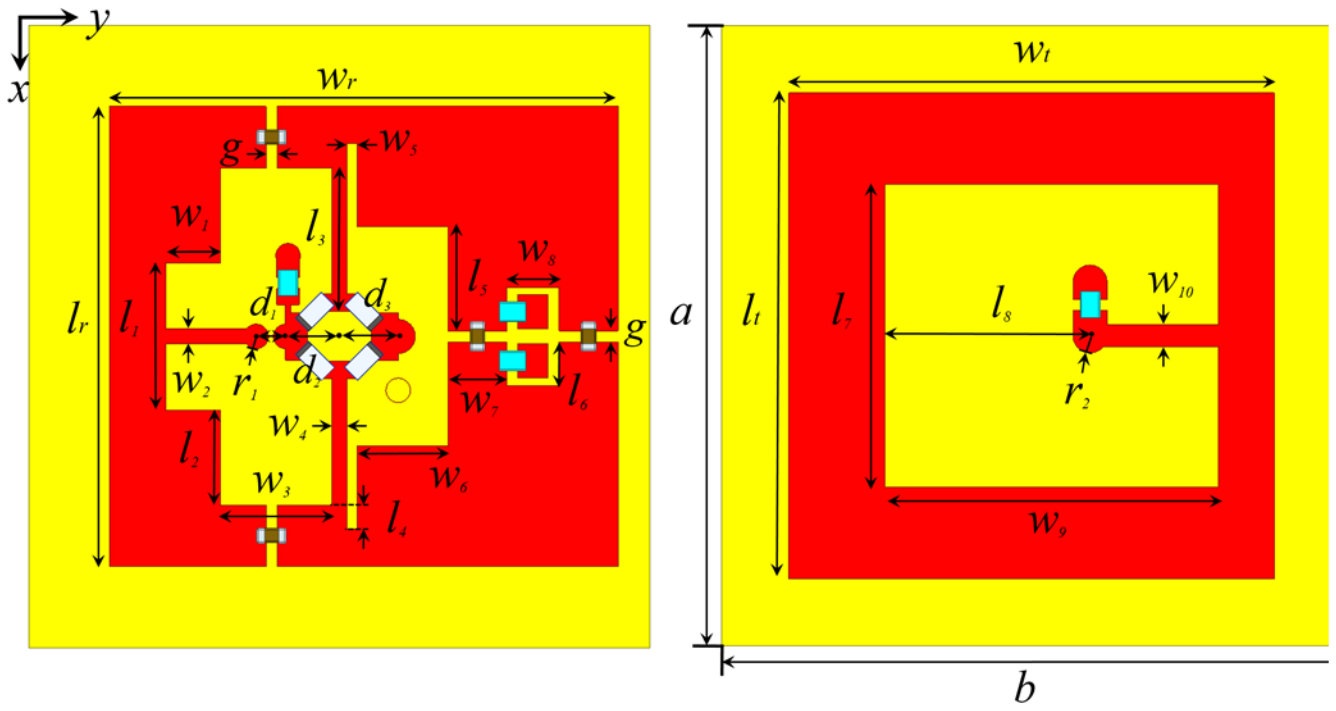

Fig. 3. Geometrical parameters of the proposed STAR-RIS unit cell. (a) Top surface. (b) Bottom surface.

TABLE I
DIMENSIONS OF THE PROPOSED STAR-RIS UNIT CELL (mm)

| $a$ | $b$ | $l_r$ | $w_r$ | $l_t$ | $w_t$ | $l_1$ | $l_2$ | $l_3$ | $l_4$ |
|---|---|---|---|---|---|---|---|---|---|
| 12.7 | 12.7 | 9.39 | 10.39 | 9.95 | 9.95 | 3.0 | 1.94 | 2.93 | 0.5 |
| $l_5$ | $l_6$ | $l_7$ | $l_8$ | $w_1$ | $w_2$ | $w_3$ | $w_4$ | $w_5$ | $w_6$ |
| 2.125 | 0.885 | 6.2 | 4.2 | 1.115 | 0.3 | 2.285 | 0.3 | 0.21 | 1.855 |
| $w_7$ | $w_8$ | $w_9$ | $w_{10}$ | $d_1$ | $d_2$ | $d_3$ | $g$ | $r_1$ | $r_2$ |
| 1.205 | 1.07 | 6.825 | 0.45 | 0.6 | 1.1 | 1.2 | 0.23 | 0.25 | 0.35 |

Four MADP-000907-14020P PIN diodes are integrated into the top radiating layer to realize the four reconfigurable states. Coilcraft 0201DS-4N9XJEW inductors are employed as RF chokes along the DC biasing paths, whereas Kyocera ATC 531Z capacitors provide DC isolation between different biasing sections while maintaining the RF performance. Three RF chokes are located on the top radiating layer and one is placed on the bottom layer. The unit-cell period is 12.7 × 12.7 mm², corresponding to approximately $0.32\lambda_0 \times 0.32\lambda_0$ at 7.6 GHz. The detailed geometrical parameters are illustrated in Fig. 3 and summarized in Table I.

The operating principle of the proposed unit cell combines orthogonal polarization conversion, switchable RF routing, and electrically controlled current reversal. The tailored top patch serves as the common receiving aperture for all four reconfigurable states. Depending on the selected switching configuration, the incident wave is directed either to the reflection or transmission path. In the reflection mode, the RF signal is routed through a pair of blind vias and the buried strip and subsequently coupled back to the top radiating aperture, generating an orthogonally polarized reflected wave. In the transmission mode, the signal is routed through the through via to the bottom ring patch, generating an orthogonally polarized transmitted wave on the opposite side of the surface. Within each mode, two mirror-symmetric switching configurations produce opposite equivalent current orientations, thereby providing two phase states with an approximately 180° phase difference. The two reflection states are denoted as states 1 and 2, while states 3 and 4 correspond to the two transmission states.

To further demonstrate the operating principle, the simulated surface current vector distributions at 7.6 GHz are presented in Fig. 4. For the reflection states in Figs. 4(a) and (b), the dominant current is mainly concentrated on the top radiating and reflection-routing structures, with only weak excitation on the bottom ring patch. In contrast, Figs. 4(c) and (d) show pronounced current excitation on the bottom ring patch, confirming that the signal is transferred to the transmission aperture through the through via. Moreover, the opposite current-vector orientations between the two states in each mode clearly demonstrate the current-reversal mechanism responsible for the approximately 180° phase difference. This mechanism enables a nearly frequency-stable phase difference and facilitates wideband 1-bit operation in both reflection and transmission modes.

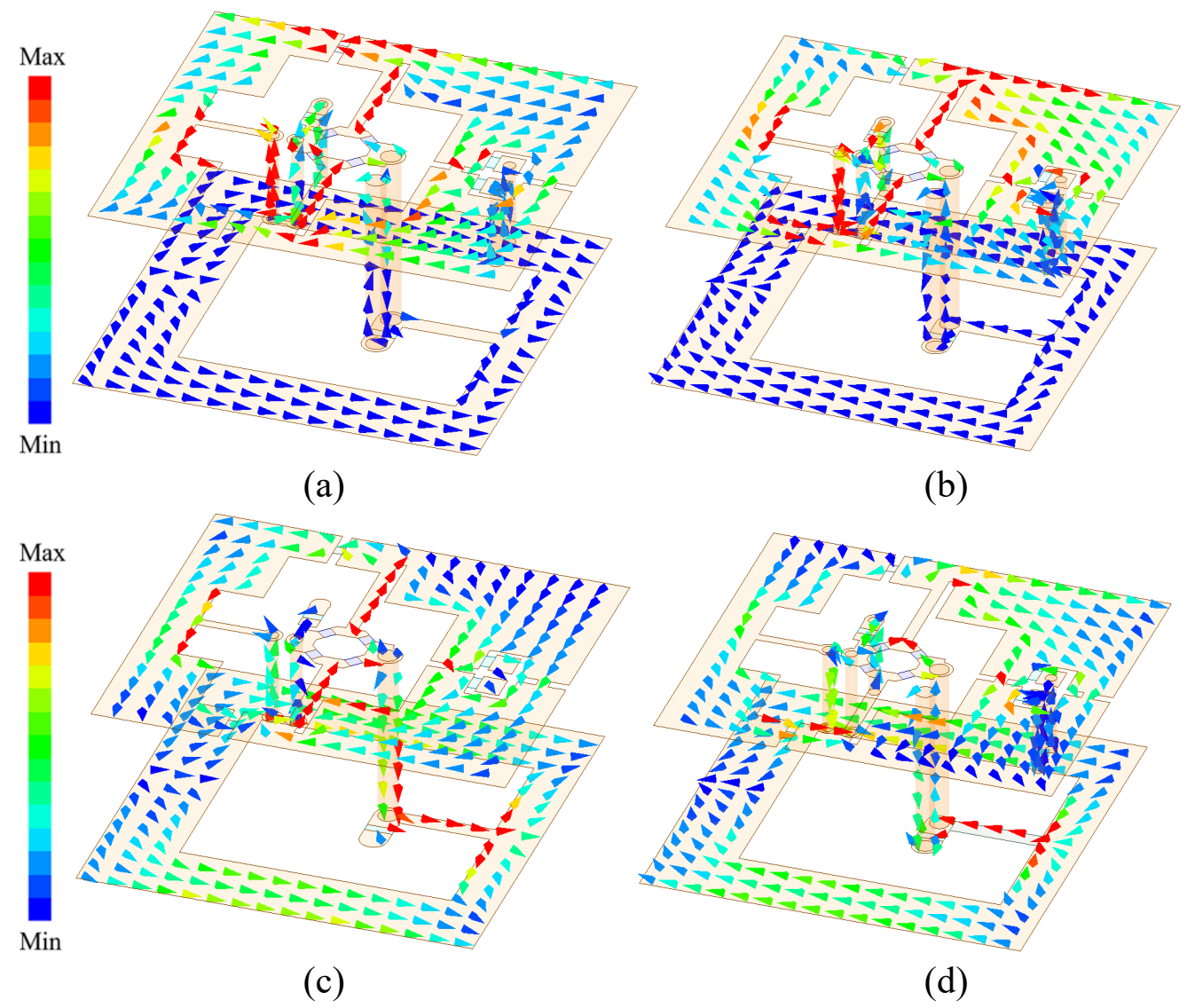

Fig. 4. Simulated surface current vector distributions at 7.6 GHz. (a) State 1 (reflection 0°). (b) State 2 (reflection 180°). (c) State 3 (transmission 0°). (d) State 4 (transmission 180°).

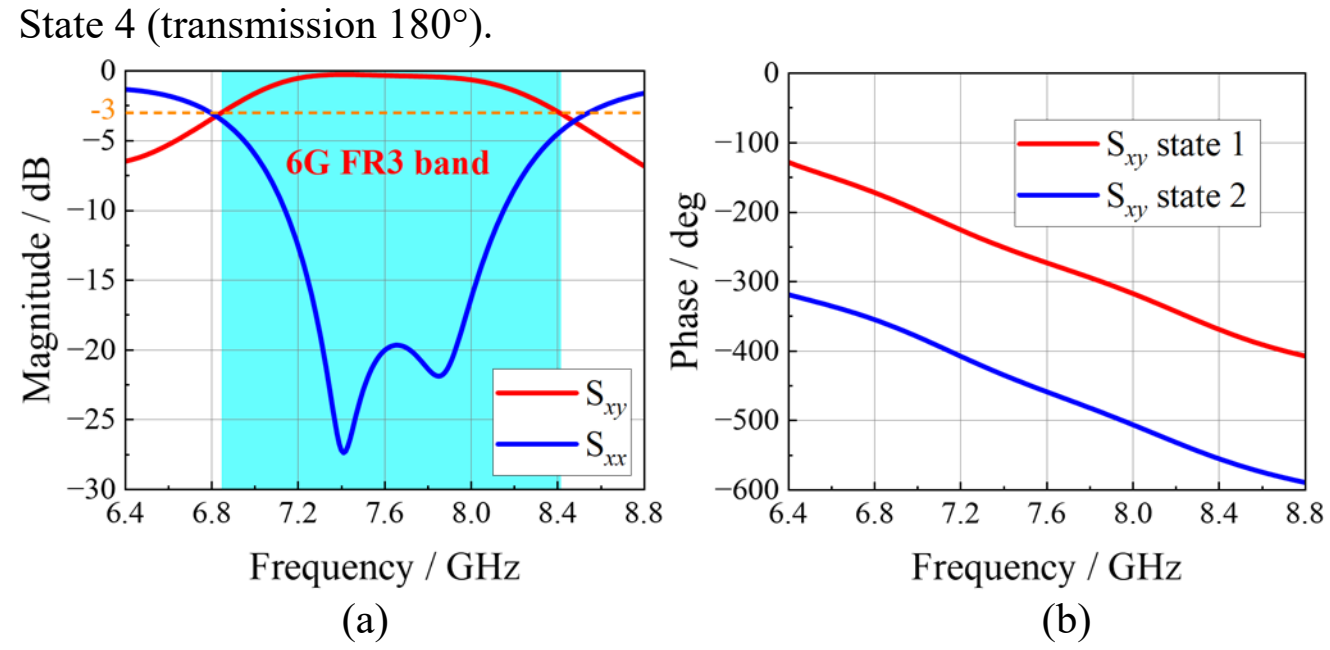

Fig. 5. Simulated reflection characteristics of the proposed STAR-RIS unit cell. (a) Magnitude. (b) Phase.

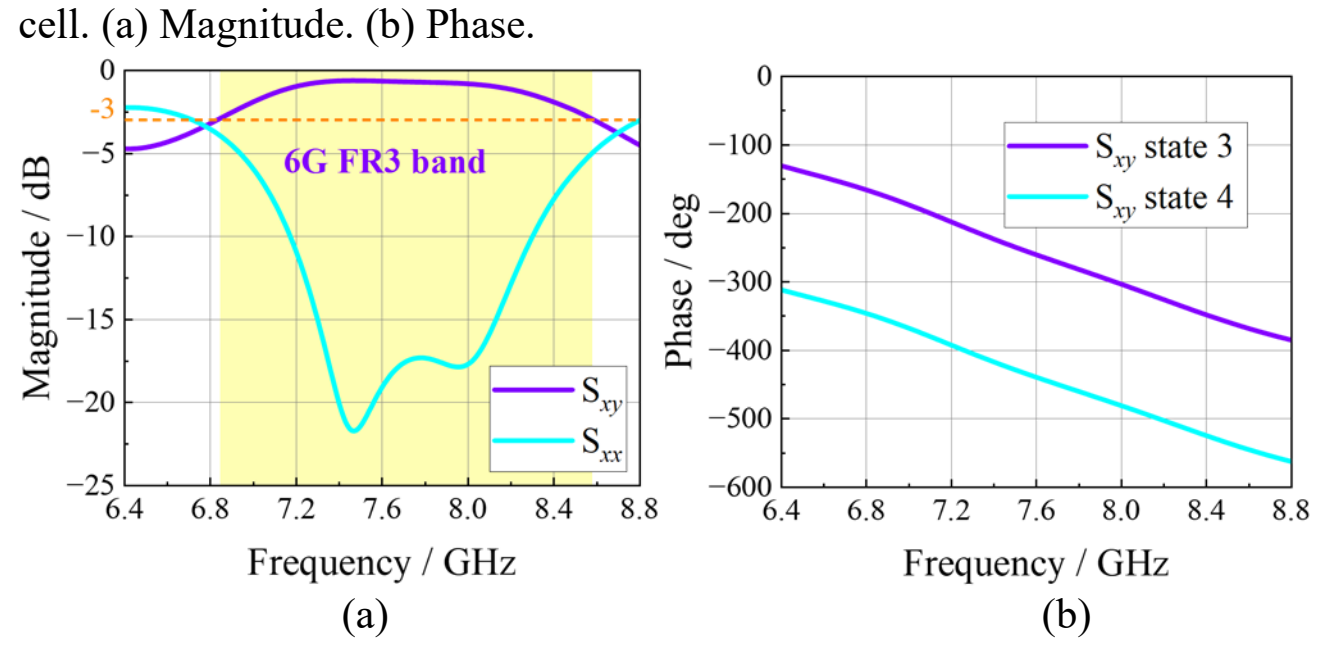

Fig. 6. Simulated transmission characteristics of the proposed STAR-RIS unit cell. (a) Magnitude. (b) Phase.

The simulated reflection response of the reflection state is presented in Fig. 5. As shown in Fig. 5(a), the reflection magnitude remains higher than −3 dB from 6.85 to 8.41 GHz, corresponding to a fractional bandwidth of 20.4% and fully covering the targeted 7.125–8.4 GHz band. More importantly, the two reflection states maintain a frequency-stable phase separation over the same frequency range. As shown in Fig. 5(b), the phase difference remains within 181°–190°, close to the ideal 180° required for 1-bit phase quantization. This wideband phase behavior is consistent with the current-reversal mechanism described above and allows the magnitude and phase-quantization requirements to be simultaneously satisfied across the targeted 6G FR3 band.

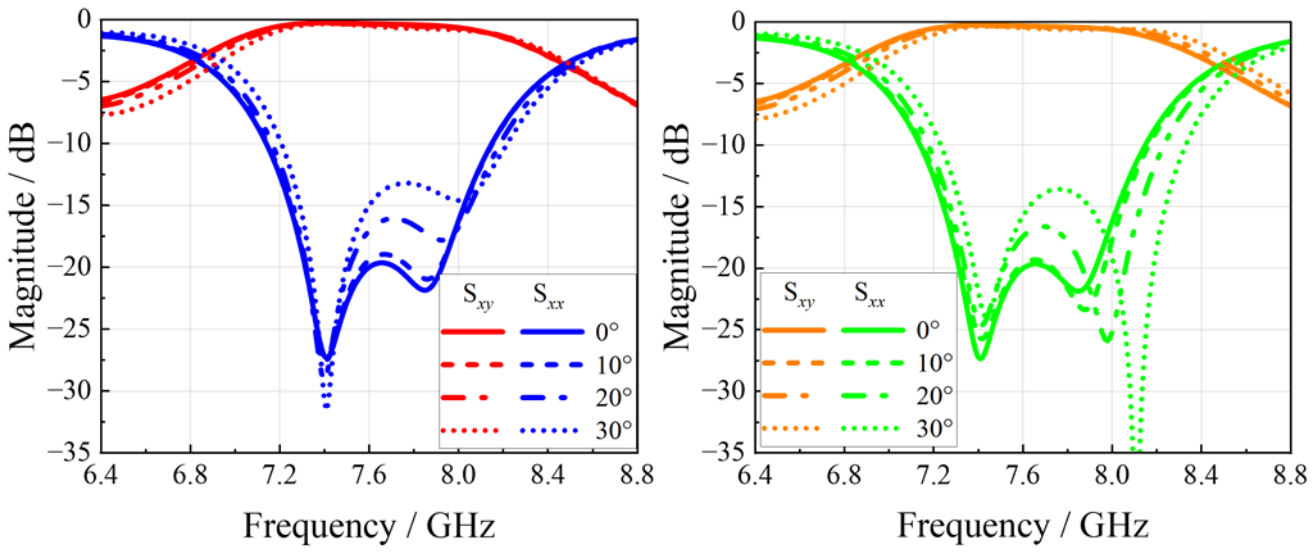

Fig. 7. Simulated angular stability of the reflection response under oblique incidence. (a) TE polarization. (b) TM polarization.

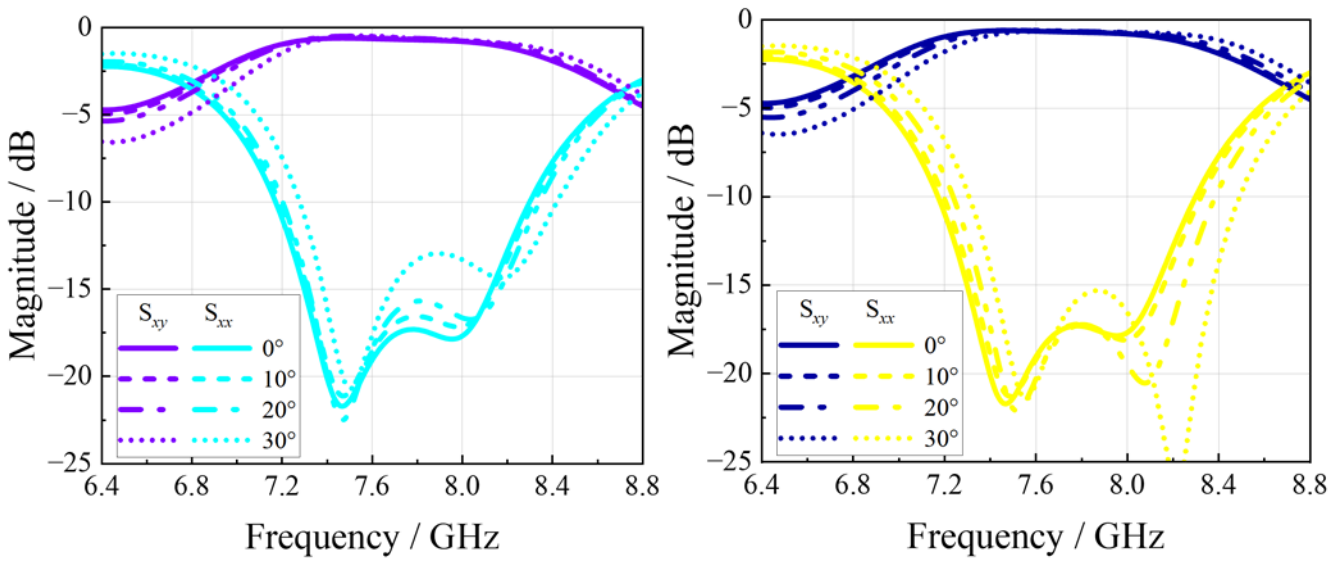

Fig. 8. Simulated angular stability of the transmission response under oblique incidence. (a) TE polarization. (b) TM polarization.

The corresponding transmission responses are presented in Fig. 6. As shown in Fig. 6(a), the transmission magnitude remains higher than −3 dB from 6.83 to 8.59 GHz, corresponding to a fractional bandwidth of 22.8%. Meanwhile, Fig. 6(b) shows that the phase difference between the two transmission states remains within 176°–181° over this frequency range, again providing a nearly 180° phase separation for wideband 1-bit quantization. Notably, the 3-dB operating bands of the reflection and transmission modes substantially overlap and both fully cover the targeted 7.125–8.4 GHz band. Therefore, the same physical aperture can provide programmable wavefront control on either side of the surface over a common wide frequency range, without requiring separate frequency-selective reflection and transmission apertures.

Finally, the angular stability of the proposed unit cell is investigated under oblique incidence. Figs. 7 and 8 show the simulated reflection and transmission responses, respectively, for both TE- and TM-polarized incident waves with incidence angles from 0° to 30°. Although slight frequency shifts and magnitude variations appear as the incident angle increases, the proposed unit cell maintains its wideband reflection and transmission characteristics over the considered angular range. In particular, stable responses are maintained up to 30° for both polarizations, indicating that the proposed unit cell can support practical STAR-RIS beamforming under non-normal illumination.

## III. STAR-RIS Design and Simulation

### A. STAR-RIS Array Design

Based on the unit cell presented in Section II, a 10 × 10 STAR-RIS array is developed, as shown in Fig. 9. The unit cells are periodically arranged with a spacing of 12.7 mm along both the *x*- and *y*- directions, resulting in an aperture size of 127 × 127 mm$^2$. The complete array preserves the multilayer configuration of the unit cell, including the top and bottom patch layers, two F4BM220 substrates, top and bottom ground planes, RO4450F prepregs, and the central FR4 layer accommodating the internal RF feeding and DC biasing network. Owing to the periodic unit-cell arrangement and distributed routing architecture, the proposed configuration supports extension to larger apertures by increasing the number of unit cells without changing the basic electromagnetic topology.

As illustrated in the exploded view of Fig. 9, the internal RF routing and DC biasing network is embedded between the two ground planes, providing EM shielding from the external radiating apertures and reducing undesired interaction between the routing network and the reflection/transmission responses. This embedded configuration also facilitates large-scale implementation, since the RF and DC routing can be periodically extended together with the array. Solder masks are applied on both outer surfaces to facilitate component soldering and provide additional protection for the fabricated PCB, while the silkscreen layers are used for component and assembly identification. The RF components required for the reconfigurable states are mounted on the corresponding top and bottom sides of the array.

To facilitate connection with the external control circuitry, multiple FFC connectors are distributed around the periphery of the STAR-RIS board. These connectors provide the electrical interface between the internal biasing network and the digital control boards, allowing the control signals to be conveniently delivered to the individual unit cells. For larger-scale implementations, additional unit cells and control channels can be incorporated by extending the periodic routing network and corresponding connector interfaces. Therefore, the proposed multilayer architecture provides a practical and scalable platform for implementing larger programmable STAR-RIS apertures.

### B. Full-Wave Beamforming Simulation

To electronically steer the reflected and transmitted waves toward desired directions, the required phase distributions over the STAR-RIS aperture are calculated by compensating the spatial phase variation of the incident field while introducing the phase gradient corresponding to the desired outgoing wave [33]–[35]. Therefore, the ideal continuous phase distribution for reflection can be expressed as [33], [7]:

$$\varphi_{ij}^{R} = k\cdot|\vec{\boldsymbol{r}}_{ij}^{e} - \vec{\boldsymbol{r}}^{f}| - k\cdot(\vec{\boldsymbol{u}}_{R}\cdot\vec{\boldsymbol{r}}_{ij}^{e}) + \Delta\varphi_{R} \qquad (1)$$

where $\varphi_{ij}^{R}$ represents the ideal continuous reflection phase of the unit cell located at $(i,j)$, $k$ is the free-space wavenumber, $\boldsymbol{r}_{ij}^{e}$ denotes the position vector of the corresponding unit cell, and $\boldsymbol{r}^{f}$ denotes the phase-center position vector of the illuminating horn. $\mathbf{u}_{R}$ is the unit vector along the desired reflected-beam direction, while $\Delta\varphi_R$ is a global phase offset used to adjust the phase reference and optimize the resulting 1-bit phase distribution. The first term in (1) compensates for the spherical phase variation introduced by the illuminating horn, whereas the second term provides the required spatial phase gradient for beam steering. The same formulation is applied to the transmission mode by replacing $\mathbf{u}_R$ and $\Delta\varphi_R$ in (1) with $\mathbf{u}_T$ and $\Delta\varphi_T$, respectively, where $\mathbf{u}_T$ denotes the unit vector along the desired transmitted-beam direction and $\Delta\varphi_T$ represents the corresponding global transmission phase offset used for the same phase-reference and quantization optimization. In this way, the same phase-synthesis framework can be applied to outgoing waves propagating toward either side of the STAR-RIS. Since two phase states are available in each operating mode, the calculated continuous phase distribution is subsequently quantized into a 1-bit phase pattern. For the reflection mode, the continuous phase values are mapped to the two reflection states presented in Section II, whereas the transmission phase distribution is mapped to the corresponding two transmission states.

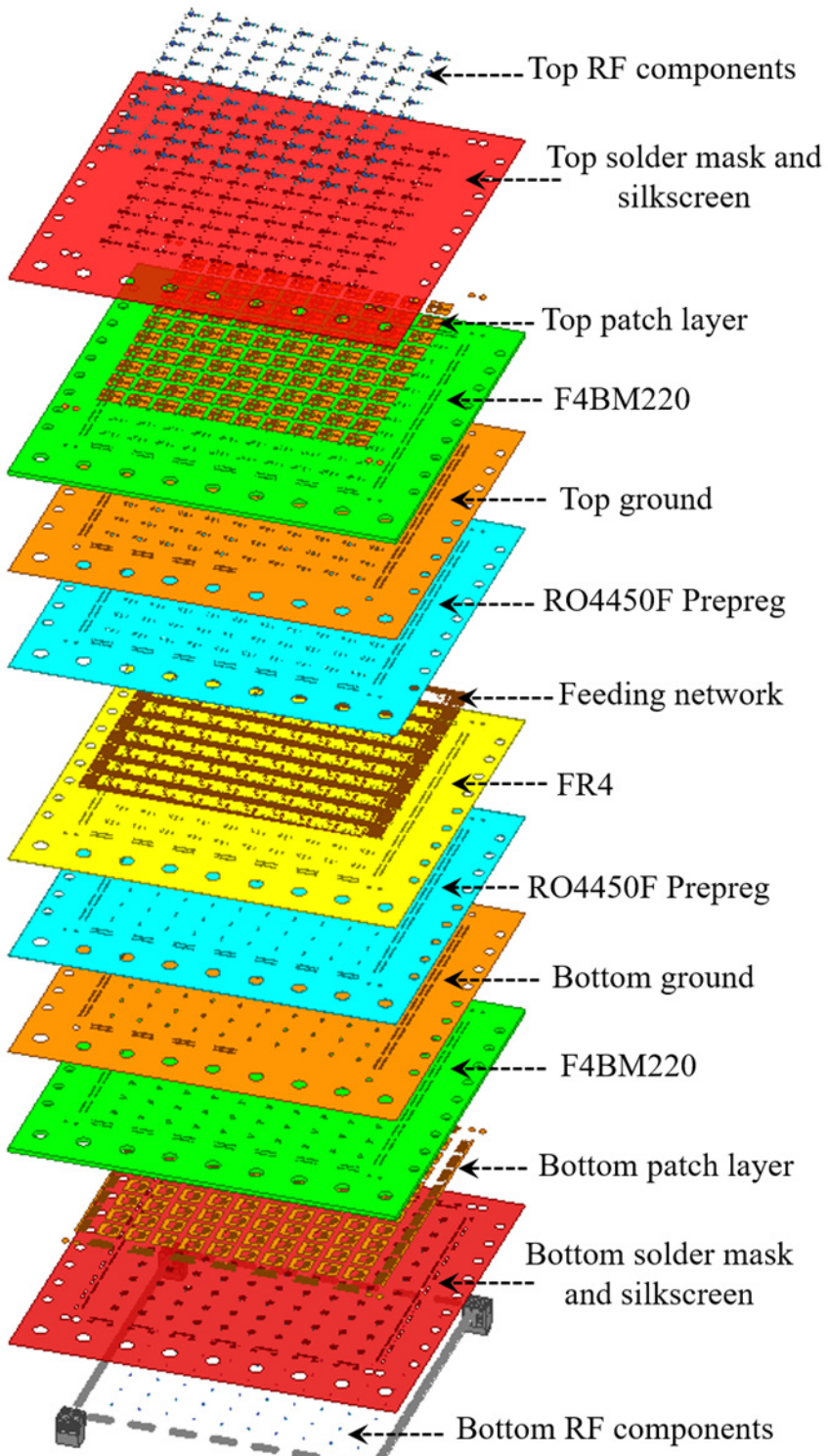

Fig. 9. Configuration and multilayer structure of the proposed STAR-RIS array.

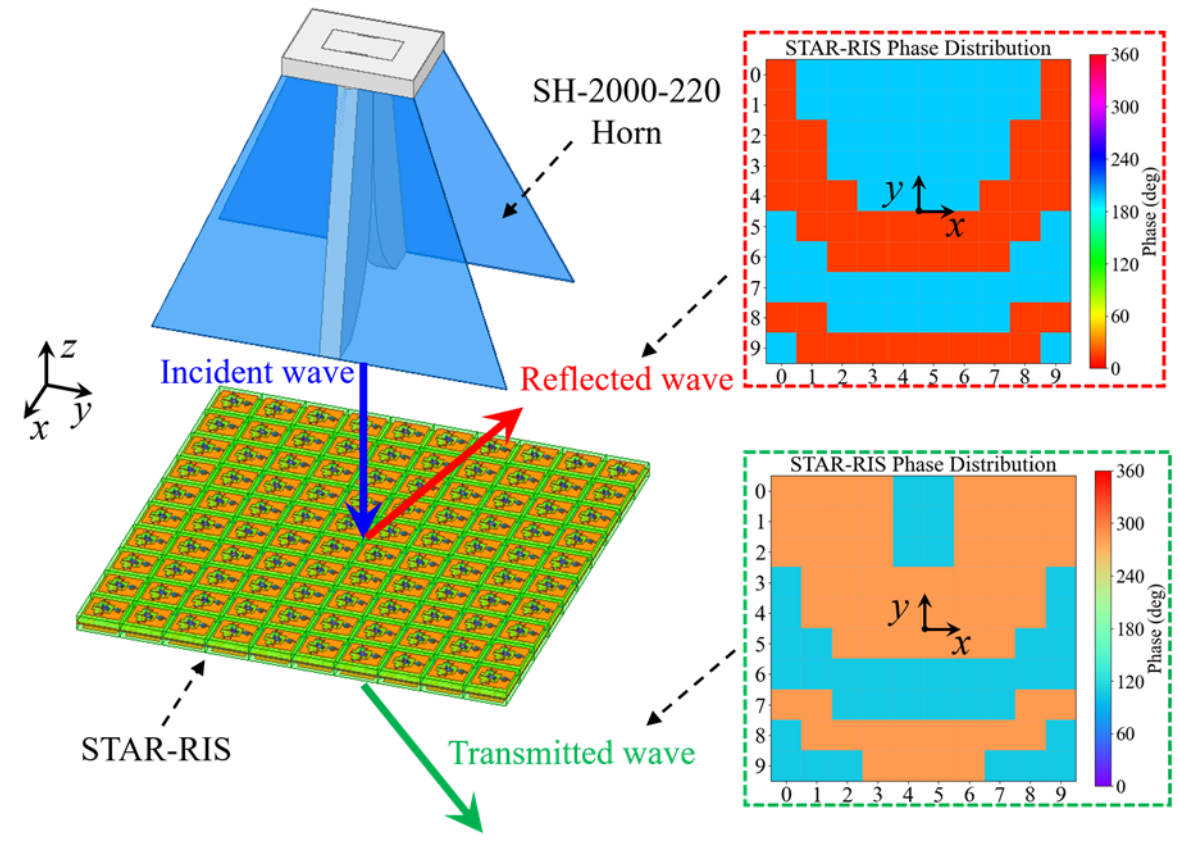

Fig. 10. Near-field full-wave simulation configuration and corresponding 1-bit phase distributions for 30° reflection and transmission

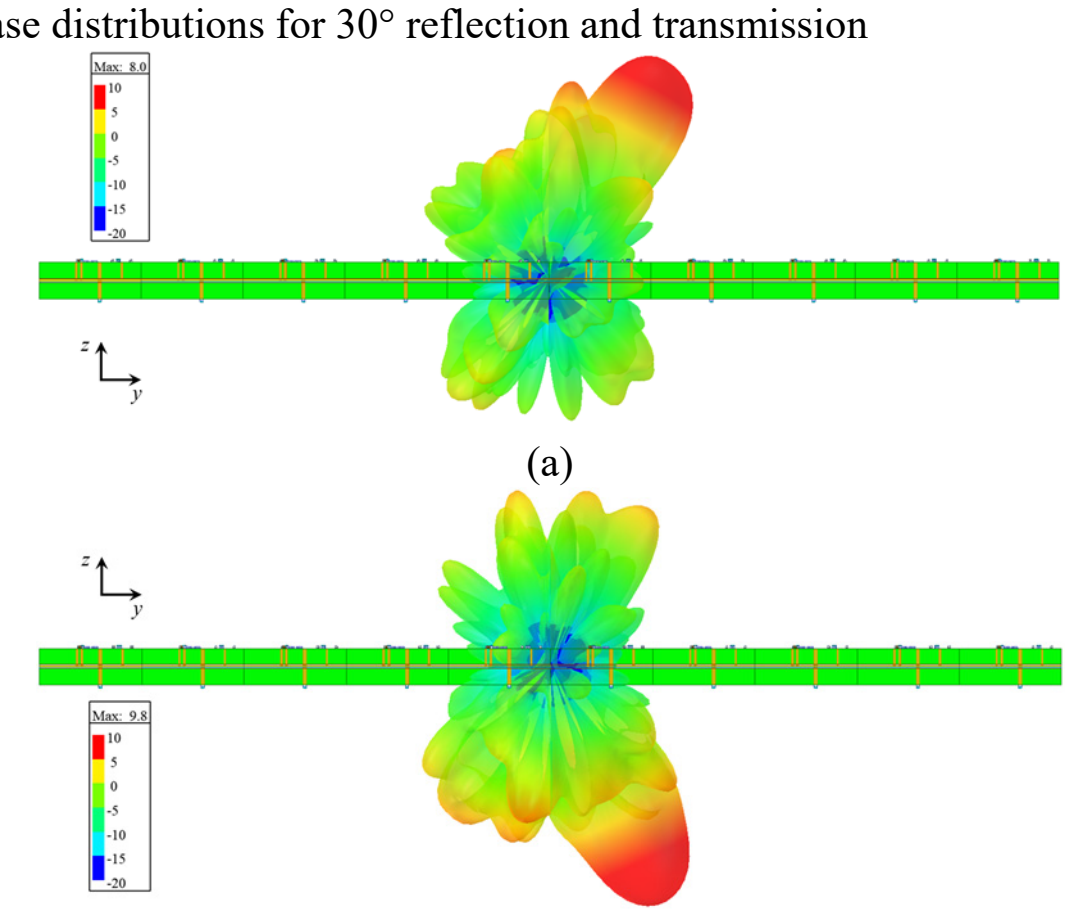

(a)

(b)

Fig. 11. Simulated 3-D radiation patterns of the proposed STAR-RIS. (a) Reflection mode. (b) Transmission mode.

For validation of the proposed STAR-RIS at the array level, near-field full-wave simulations are conducted using the complete 10 × 10 structure. As shown in Fig. 10, an MVG SH-2000-220 standard horn is employed as the illumination source, with its phase center located approximately 80 mm from the center of the STAR-RIS. The incident angle is set to 0°, while the desired reflection and transmission angles are both selected as 30°. The calculated continuous phase distributions are then quantized according to the two available states in each operating mode, and the corresponding 1-bit reflection and transmission phase patterns are also shown in Fig. 10. For the reflection case, the unit cells are configured using the two reflection states such that the incident wave is redirected toward the desired 30° direction in the same hemisphere as the illumination source. For the transmission case, the unit cells are configured using the two transmission states, allowing the incident wave to propagate through the STAR-RIS and form a directive beam toward 30° in the opposite hemisphere.

The corresponding simulated three-dimensional radiation patterns are presented in Fig. 11. Clear directive beams are obtained toward the desired directions for both operating modes, demonstrating effective wavefront control on both sides of the STAR-RIS. The simulated realized gains are 8.0 dBi for the reflection mode and 9.8 dBi for the transmission mode. These results confirm that the unit-cell characteristics presented in Section II can be successfully extended to array-level beamforming. It should be noted that the desired beam direction is not fixed by the physical structure of the STAR-RIS. Once the illumination wave and target outgoing direction are specified, the corresponding phase distribution can be recalculated using (1) with the appropriate propagation vector and phase offset, followed by 1-bit phase quantization. Therefore, different reflected or transmitted beam directions can be electronically generated without modifying the physical aperture.

## C. Control Circuit Design

To implement the calculated phase distributions on the practical STAR-RIS hardware, a programmable digital control system is developed, as shown in Figs. 12 and 13. As shown in Fig. 12, the 74HC595 shift registers are employed to convert the serial digital data generated by the ESP-WROOM-32 module into parallel control outputs. By cascading multiple shift registers, a large number of STAR-RIS control channels can be addressed using only a limited number of digital control interfaces, which also facilitates extension to larger arrays. The SN74ACT244 logic buffers are inserted between the shift-register outputs and the STAR-RIS biasing network to provide logic buffering and sufficient driving capability. LEDs integrated on the buffer board provide a direct visual indication of the applied digital states.

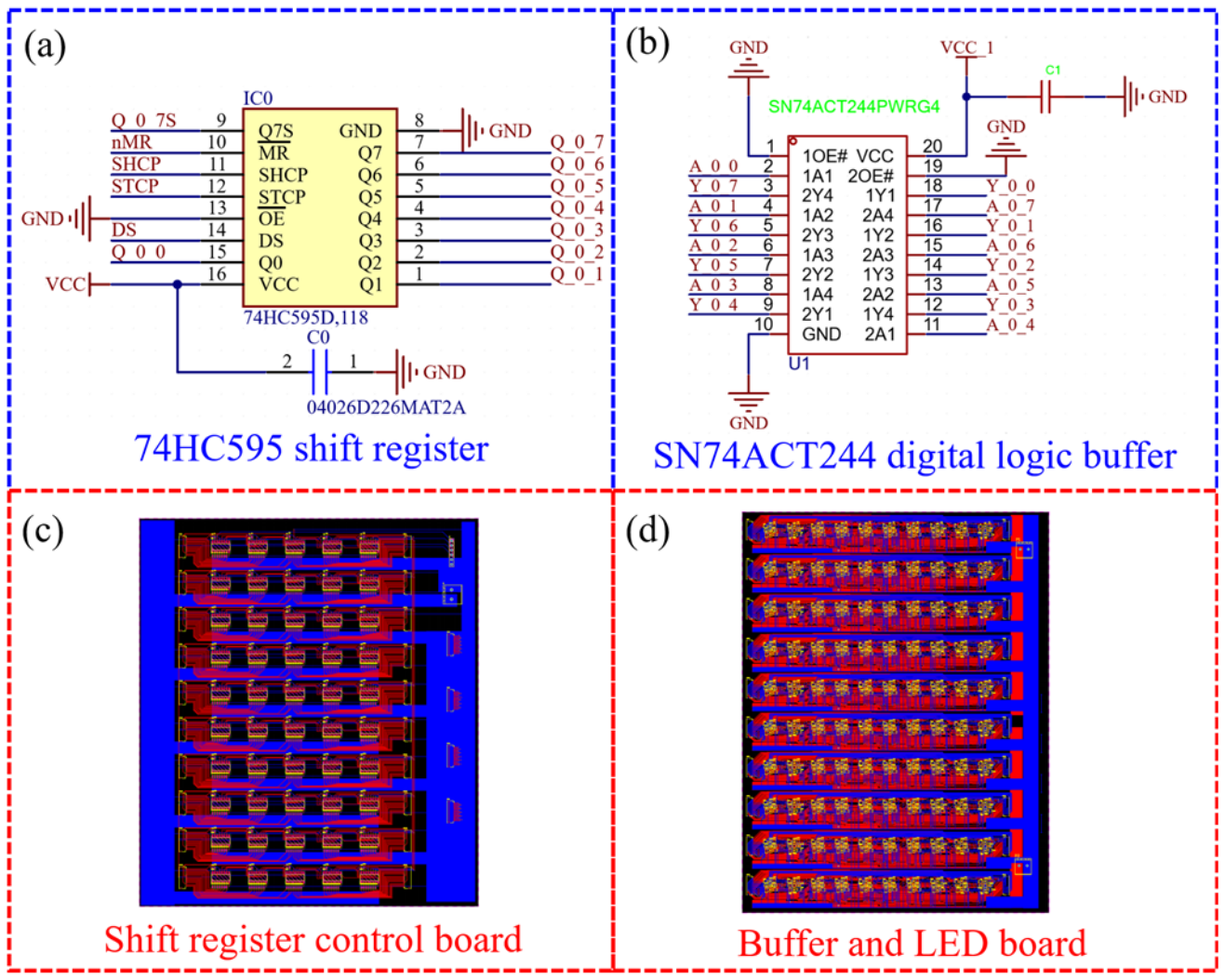

Fig. 12. Programmable control circuitry of the proposed STAR-RIS. (a) 74HC595 shift-register stage. (b) SN74ACT244 logic-buffer stage. (c) Shift-register control-board layout. (d) Buffer and LED board layout.

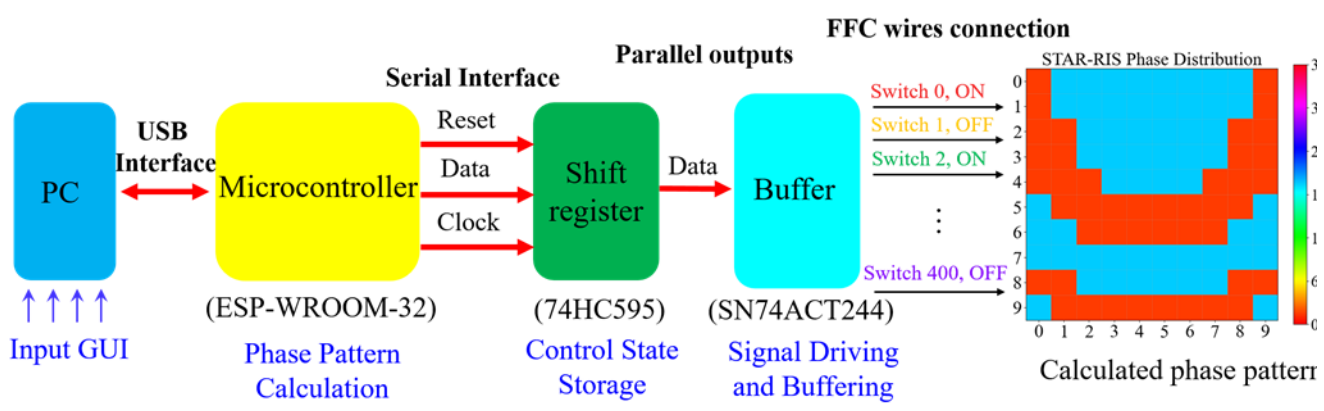

Fig. 13. Block diagram of the STAR-RIS digital control architecture and pattern-generation flow.

The overall control procedure is summarized in Fig. 13. The operating frequency, reflection or transmission mode, illumination geometry, and desired beam direction are first specified through the input GUI. Based on these parameters, the required phase distribution is calculated using (1) with the corresponding reflection or transmission propagation vector and phase offset. The continuous phase distribution is then converted into the corresponding 1-bit pattern and mapped to the available reflection or transmission states. The resulting digital control sequence is transferred from the ESP-WROOM-32 module to the cascaded 74HC595 shift registers through the serial data and clock interfaces. After serial-to-parallel conversion, the control signals are passed through the SN74ACT244 buffers and delivered to the STAR-RIS through the peripheral connectors and flexible flat cables.

## IV. Fabrication and Measurement

### A. Fabricated Prototypes

To experimentally validate the proposed STAR-RIS, the complete 10 × 10 array together with the corresponding digital control circuits is fabricated, as shown in Fig. 14. Fig. 14(a) presents the front and back views of the fabricated STAR-RIS array. The RF components and DC biasing networks described in Sections II and III are fully integrated into the multilayer PCB, while the FFC connectors located around the periphery of the array provide interfaces to the external control circuitry.

The fabricated control system consists of two separate circuit boards, as shown in Figs. 14(b) and (c). The shift-register control board integrates the ESP-WROOM-32 microcontroller, cascaded 74HC595 shift registers, FFC connectors, and DC power interfaces. The microcontroller receives the calculated digital pattern from the PC and transfers the corresponding serial control sequence to the cascaded shift registers, which generate the parallel control signals required by the STAR-RIS. The bottom side of the board accommodates the distributed power and ground routing for the control circuitry.

Fig. 14(c) shows the fabricated buffer and LED board, which incorporates the SN74ACT244 digital logic buffers, status-indication LEDs, FFC connectors, and DC power interfaces. The buffer stages provide the driving capability required for controlling the PIN diodes, while the LEDs provide a direct visual indication of the applied digital states during programming and measurement. The shift-register board, buffer board, and STAR-RIS array are interconnected through FFC cables, forming a complete programmable STAR-RIS hardware platform for the experimental characterization presented in the following subsections.

### B. RIS Characterization

Before evaluating the beamforming performance of the complete STAR-RIS, the practical magnitude and phase responses of the fabricated array are first experimentally characterized. Following the two-horn characterization approach commonly adopted for RIS [36], [37], [7], all unit cells are programmed to an identical state so that the measured response represents the corresponding uniform aperture state. The same horn antennas used in the subsequent near-field measurements are employed for this characterization, where an MVG SH-2000-220 horn serves as the transmitting antenna and an MVG SH4000-068 horn is used as the receiving antenna. For the reflection characterization, the Tx and Rx horns are positioned symmetrically with respect to the surface normal of the STAR-RIS, whereas for the transmission characterization, the receiving horn is placed on the opposite side of the surface. By sequentially switching the complete array between the two reflection states and the two transmission states, the practical magnitude responses and phase differences of the fabricated STAR-RIS can be extracted over frequency.

Fig. 15 compares the simulated and measured reflection characteristics of the proposed STAR-RIS. The measured normalized reflection magnitude maintains a 3-dB bandwidth from 6.86 to 8.37 GHz, corresponding to a fractional bandwidth of 19.8%. This measured bandwidth agrees well with the simulated wideband reflection response presented in Section II. The measured phase responses of reflection states 1 and 2 also follow the simulated frequency-dependent trends. Over the observation range of 6.4 – 8.8 GHz, the measured phase difference between the two reflection states varies from approximately 150° to 195°. The agreement between simulation and measurement verifies that the electrically controlled current-reversal mechanism introduced in Section II is preserved after array fabrication and integration. The corresponding transmission characteristics are shown in Fig. 16. The measured normalized transmission magnitude exhibits a 3-dB bandwidth from 6.69 to 8.34 GHz, corresponding to a fractional bandwidth of 21.9%. Similar to the reflection case, the measured transmission response follows the overall trend predicted by the full-wave simulation. The measured phase responses of transmission states 3 and 4 exhibit a phase difference ranging from approximately 134° to 176° over 6.4–8.8 GHz. Although the measured phase separation deviates more noticeably from the ideal 180° value at some frequencies, the two transmission states still maintain a clear and relatively stable phase separation over a broad frequency range, enabling practical 1-bit transmission beamforming. The discrepancies can be mainly attributed to practical fabrication tolerances, component parasitics, and substrate property variations.

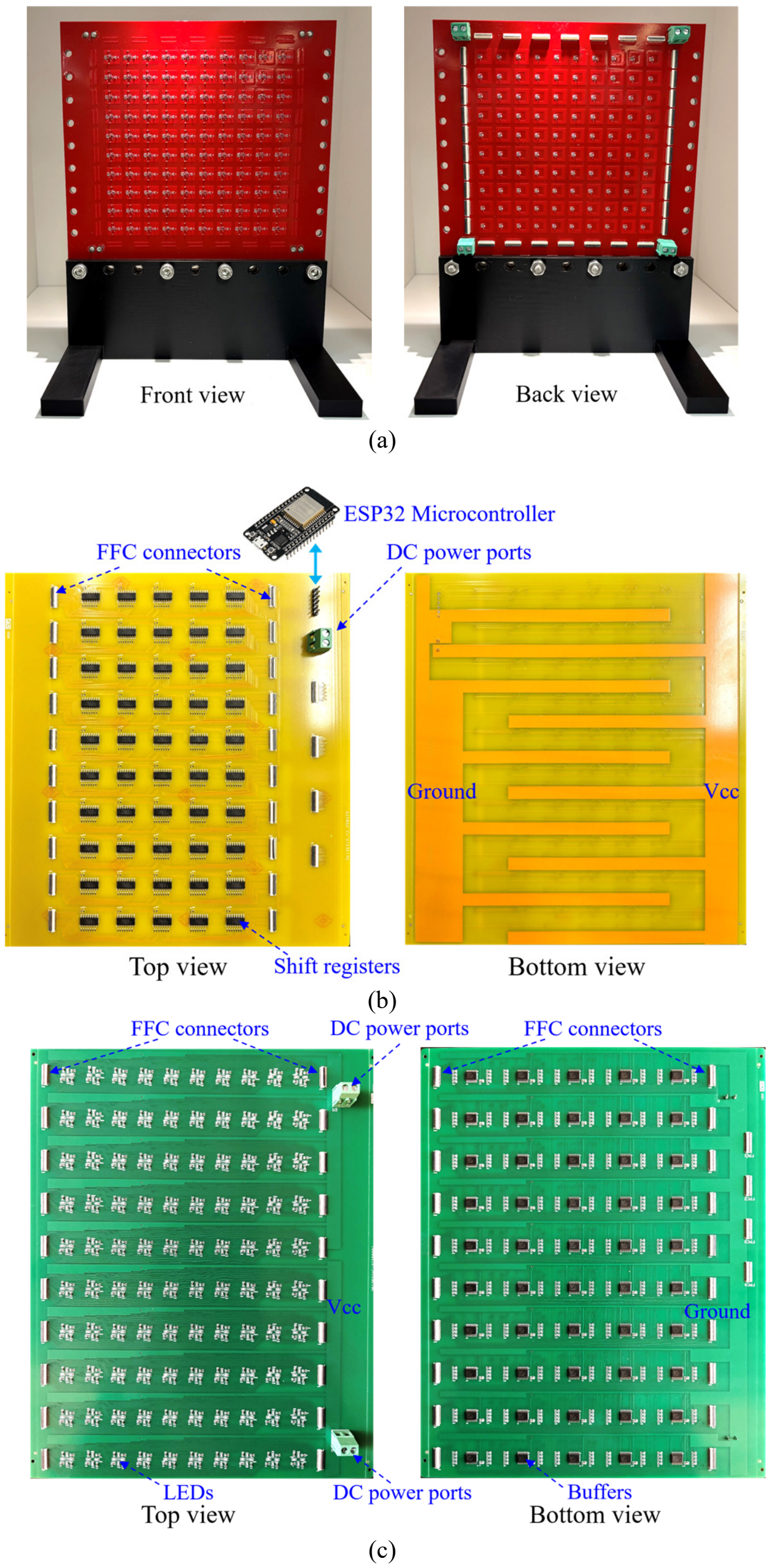

Fig. 14. Fabricated prototypes of the proposed STAR-RIS system. (a) 10 × 10 STAR-RIS array. (b) Shift-register control board. (c) Buffer and LED board.

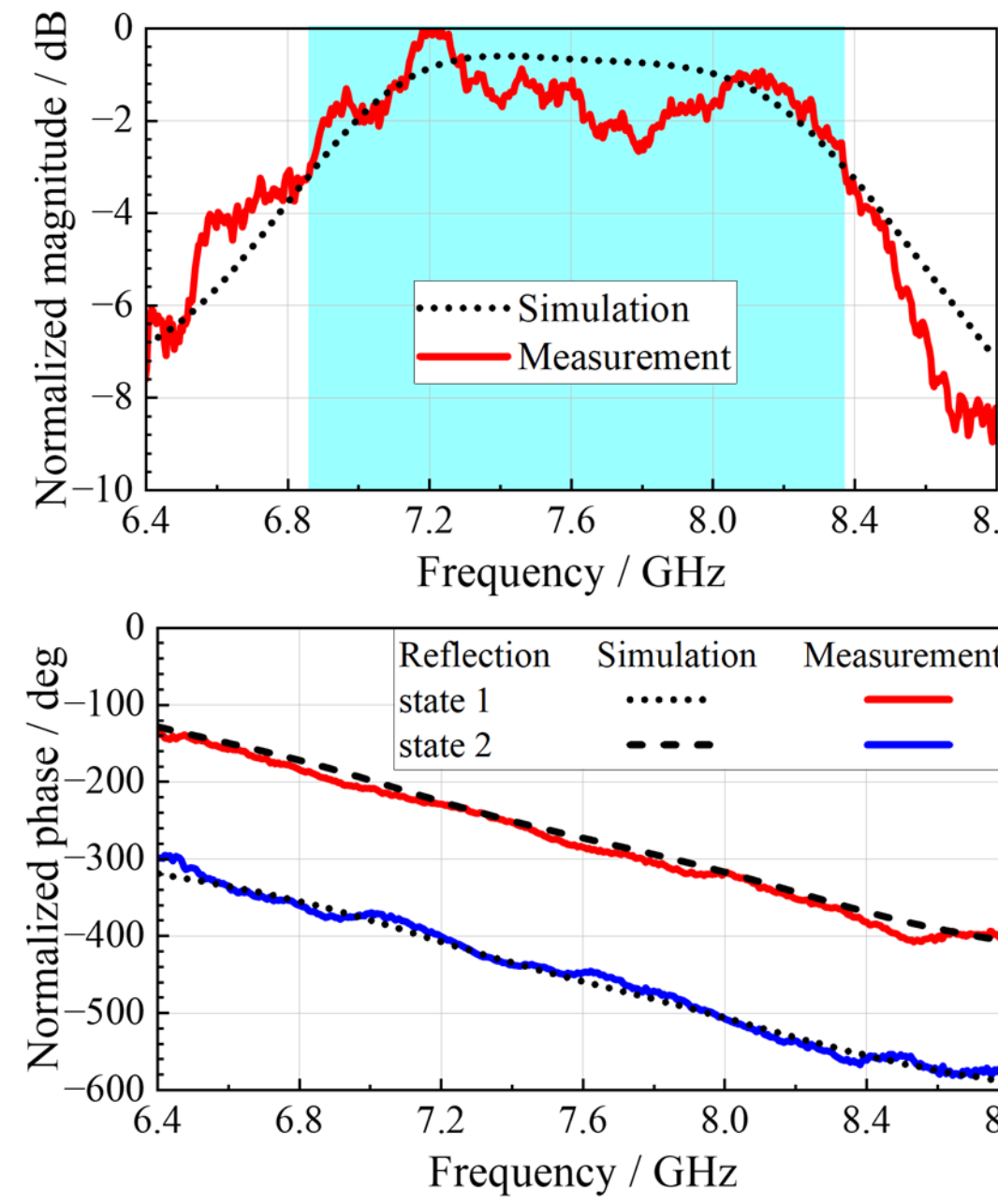

Fig. 15. Comparison of measured and simulated normalized reflection characteristics of the fabricated STAR-RIS. (a) Magnitude. (b) Phase.

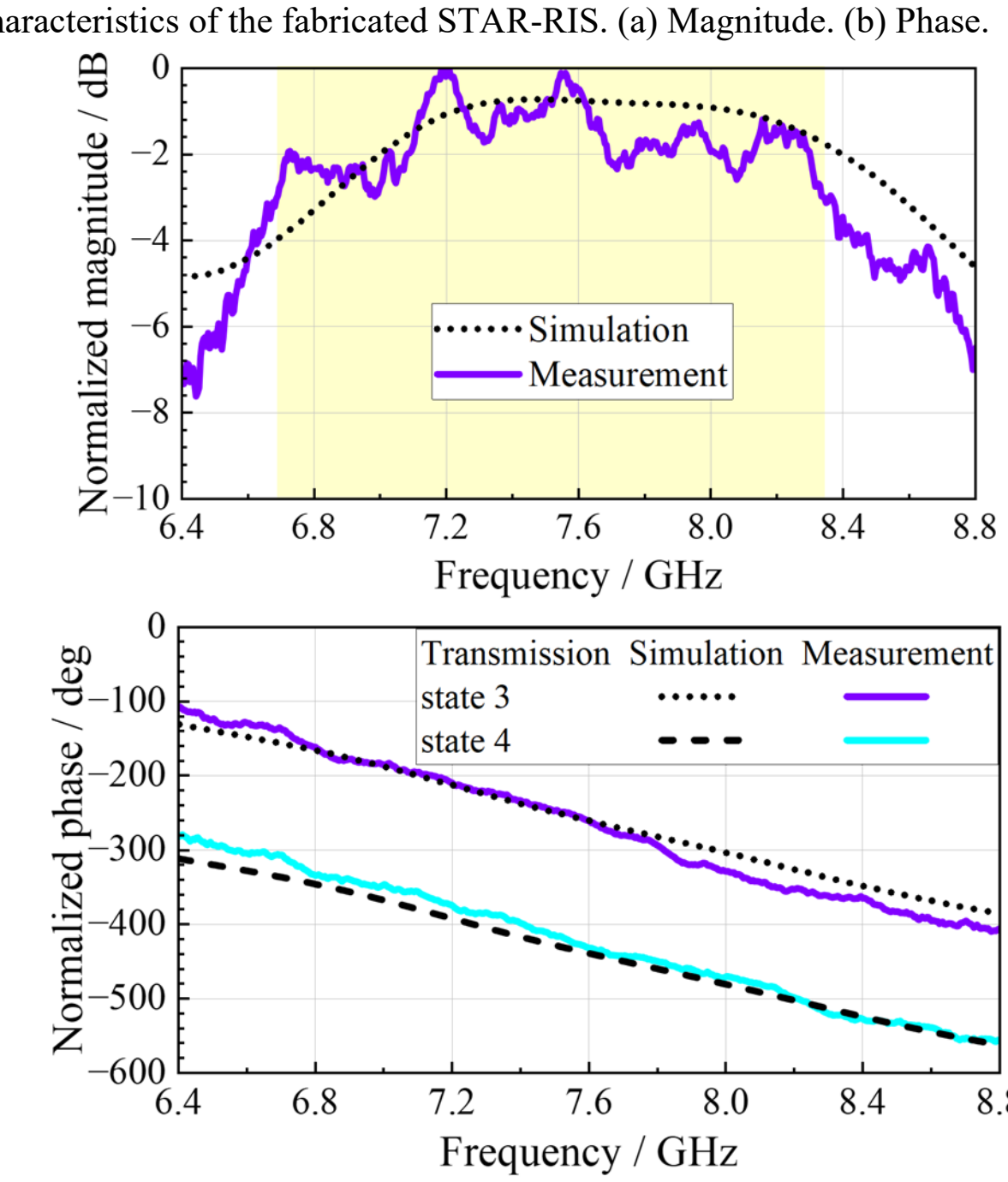

Fig. 16. Comparison of measured and simulated normalized transmission characteristics of the fabricated STAR-RIS. (a) Magnitude. (b) Phase.

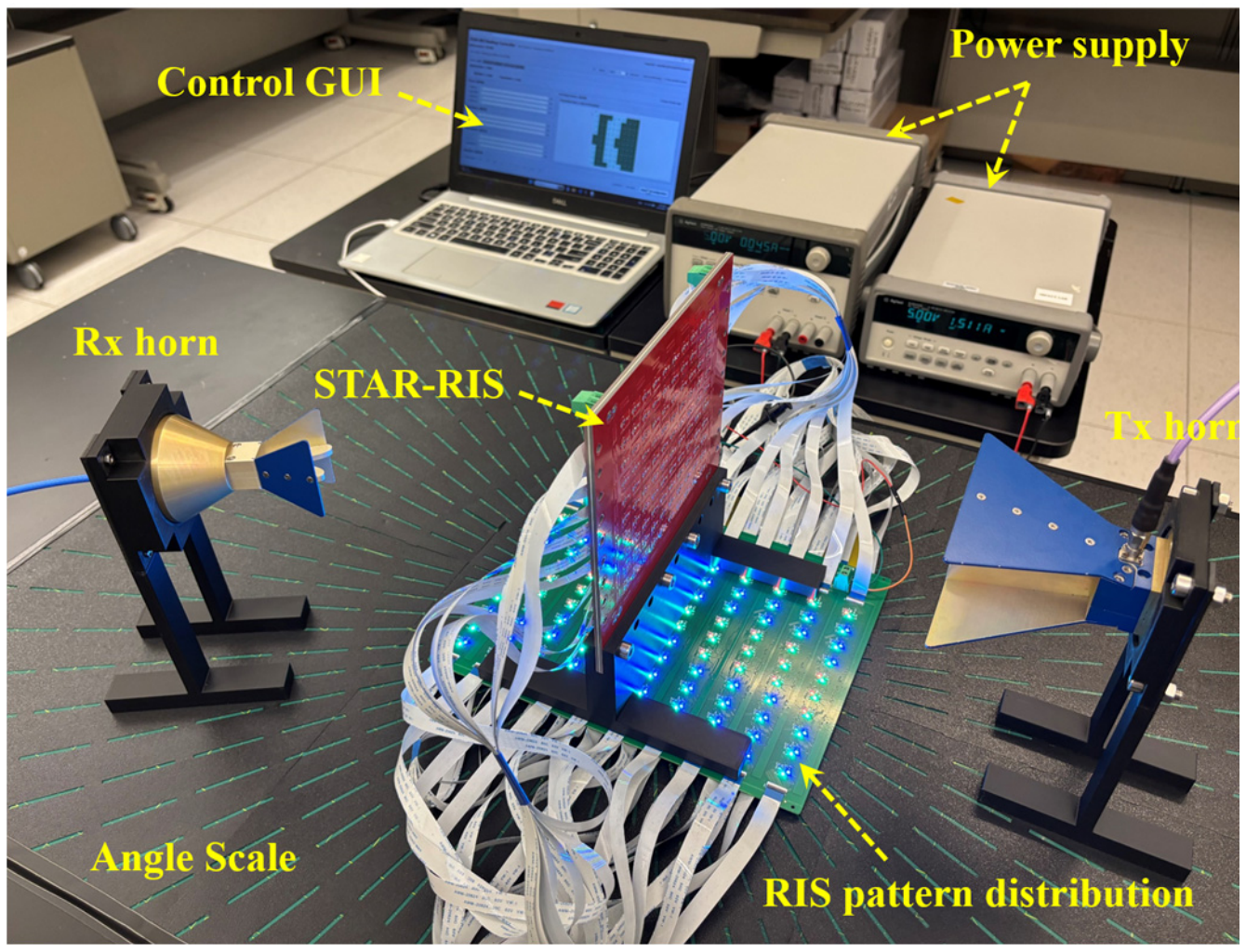

Fig. 17. Near-field measurement setup.

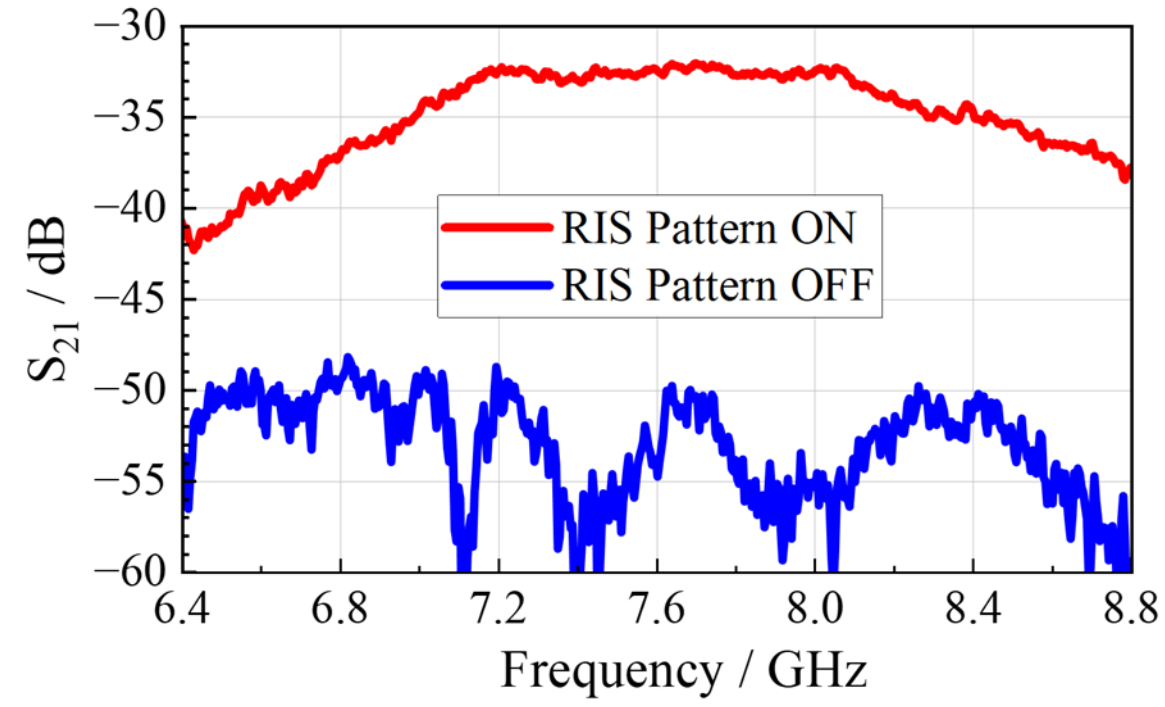

Fig. 18. Measured near-field $S_{21}$ versus frequency under RIS ON and OFF conditions for the reflection mode.

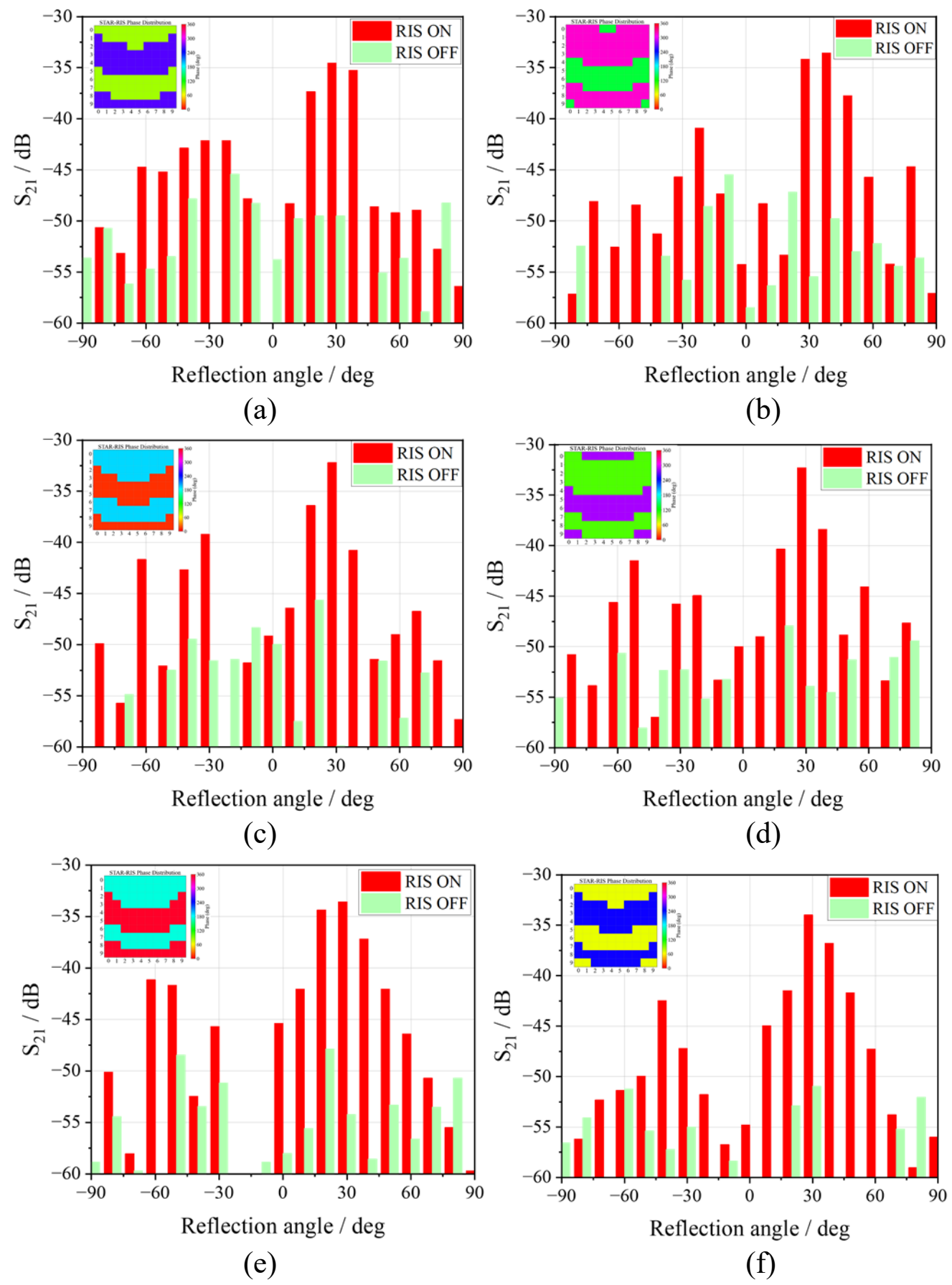

Fig. 19. Measured near-field reflection patterns for a 30° target beam at (a) 6.8 GHz, (b) 7.1 GHz, (c) 7.4 GHz, (d) 7.8 GHz, (e) 8.1 GHz, and (f) 8.4 GHz.

### C. Near-Field Measurement

To characterize the practical beamforming performance of the proposed STAR-RIS, near-field measurements are first conducted for the reflection scenario. The measurement setup is shown in Fig. 17. An MVG SH-2000-220 horn is employed as the transmitting antenna to illuminate the STAR-RIS, while an MVG SH4000-068 horn is used as the receiving antenna. The Tx horn is positioned approximately 20 cm from the STAR-RIS with normal incidence, whereas the Rx horn is located approximately 25 cm from the surface and can be rotated along the angular scale to characterize the reflected field over different observation angles. The relatively short Tx–RIS and RIS–Rx distances are selected to characterize the near-field beamforming performance of the fabricated STAR-RIS.

The wideband reflection performance is first investigated by comparing the received $S_{21}$ when the calculated RIS pattern is applied with that obtained in the RIS OFF condition. As shown in Fig. 18, a clear received-signal enhancement is maintained over a broad frequency range. From 6.8 to 8.4 GHz, the proposed STAR-RIS provides an average measured enhancement of approximately 19.8 dB, demonstrating that the wideband magnitude and phase responses characterized in Section IV-B can be effectively translated into practical array-level beamforming performance. In particular, the relatively stable ON-state response over this frequency range confirms that the proposed STAR-RIS can maintain effective reflection beamforming across the targeted wide operating band.

To further verify the beamforming performance over frequency, the reflected field is measured at six representative frequency points of 6.8, 7.1, 7.4, 7.8, 8.1, and 8.4 GHz, as shown in Fig. 19. For all cases, the incident angle is fixed at 0°, while the desired reflection angle is configured to 30°. The corresponding 1-bit phase distributions are recalculated according to the operating frequency and applied to the fabricated STAR-RIS. Clear reflected beams are observed close to the desired 30° direction throughout the entire frequency range. The measured $S_{21}$ values at the desired beam direction are −34.6, −33.5, −32.1, −32.3, −33.6, and −33.9 dB at 6.8, 7.1, 7.4, 7.8, 8.1, and 8.4 GHz, respectively, showing a variation of only 2.5 dB over the 1.6 GHz measurement bandwidth. The corresponding highest sidelobe levels are −42.1, −40.9, −39.2, −41.5, −41.1, and −42.5 dB, providing approximately 7.1–9.2 dB separation from the desired beams. In comparison, the received levels under the RIS OFF condition are −49.5, −55.4, −59.9, −53.9, −54.2, and −50.9 dB, respectively. These results verify that effective reflection beamforming is maintained over the wide frequency range. It is also noted that relatively low measured levels appear around the −10° to 10° region in Fig. 18. This is mainly caused by the physical blockage of the receiving path by the illumination horn rather than by the radiation characteristics of the STAR-RIS itself.

The beamscanning capability is subsequently evaluated at the center frequency of 7.6 GHz. As shown in Fig. 20, the desired reflection direction is electronically reconfigured to −30°, −15°, 15°, and 30° by changing only the applied 1-bit phase distribution while maintaining normal incidence. Well-defined reflected beams are obtained at all four target directions. The measured $S_{21}$ values at the desired beam directions are −32.4, −32.6, −33.0, and −32.2 dB, respectively, corresponding to a maximum variation of only 0.8 dB across the considered scanning range. Meanwhile, the highest sidelobe levels are −42.3, −41.5, −42.6, and −40.7 dB, corresponding to approximately 8.5–9.9 dB separation from the main beams. These results experimentally confirm the programmable reflection beam-steering capability of the proposed STAR-RIS over the angular range of ±30°.

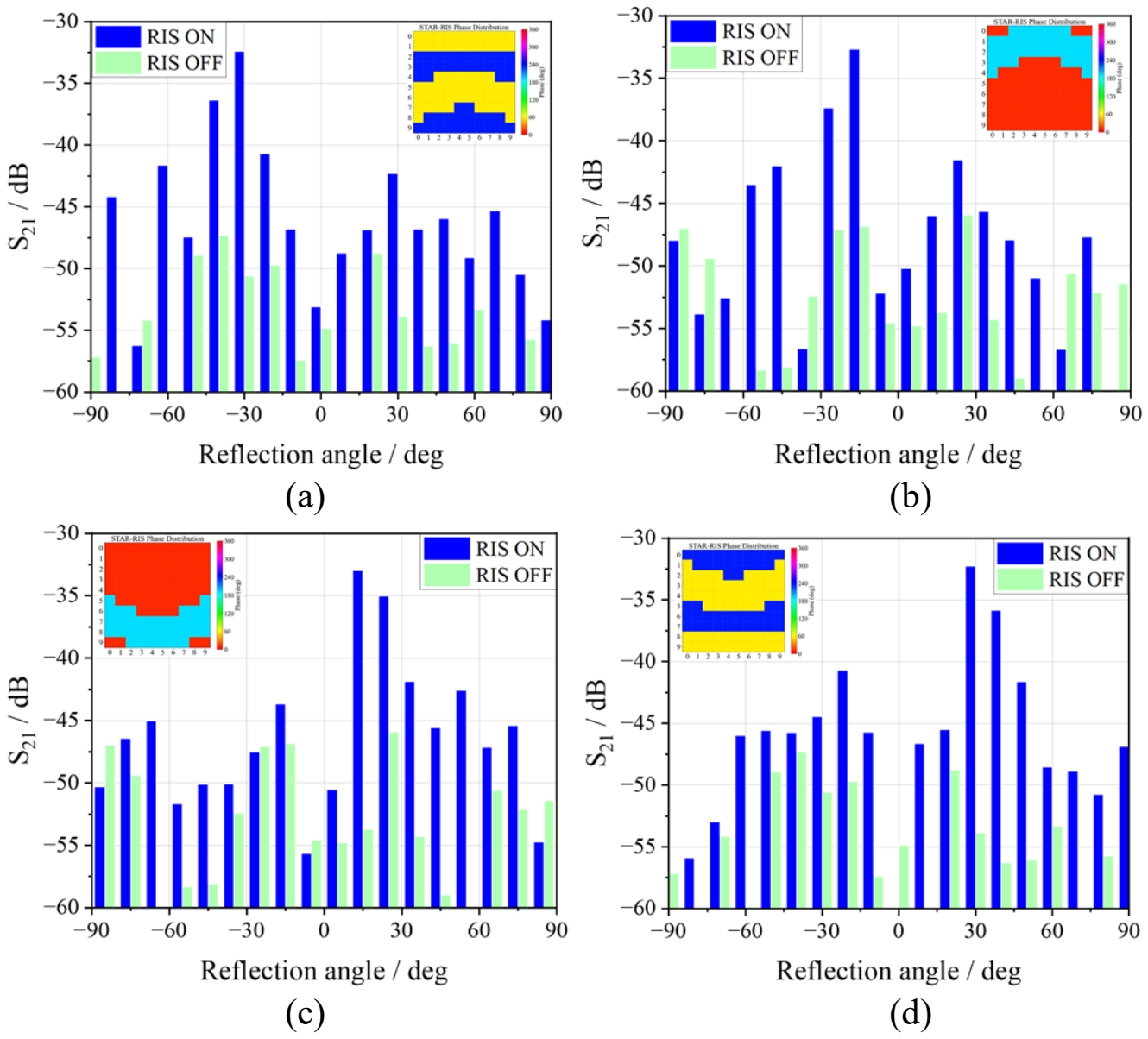

Fig. 20. Measured near-field reflection beam scanning at 7.6 GHz for target angles of (a) −30°, (b) −15°, (c) 15°, and (d) 30°.

For the transmission measurement, the overall setup follows the reflection case, with the main difference being the position of the Rx horn. The Tx horn and STAR-RIS remain unchanged, while the Rx horn is moved to the opposite side of the STAR-RIS to characterize the transmitted field. During the measurement, the Rx horn is rotated along the angular scale on the transmission side, enabling the transmitted beam to be recorded over different observation angles.

The broadband behavior of the transmission mode is examined next through the RIS pattern ON/OFF comparison shown in Fig. 21. A pronounced separation between the two conditions is maintained across the operating band, with an average received-signal enhancement of approximately 22.6 dB from 6.8 to 8.4 GHz. Together with the reflection results presented above, this broadband improvement demonstrates that the fabricated STAR-RIS can provide effective wavefront manipulation on both sides of the aperture over a common wide frequency range.

The frequency robustness of the transmitted beam is further characterized at 6.8, 7.1, 7.4, 7.8, 8.1, and 8.4 GHz, as presented in Fig. 22. With normal incidence and a desired transmission angle of 30°, the corresponding 1-bit transmission pattern is updated for each frequency. The measured $S_{21}$ levels at the desired direction are −31.6, −31.1, −30.3, −29.9, −29.8, and −32.2 dB, respectively, resulting in a variation of 2.4 dB across the 1.6 GHz measurement bandwidth. The corresponding sidelobe levels are −39.2, −37.7, −36.9, −38.1, −39.7, and −41.1 dB, which are approximately 6.6–9.9 dB lower than the desired beam levels. In contrast, the RIS OFF responses are significantly weaker, with measured levels of −51.7, −52.2, −50.7, −54.5, −54.8, and −51.7 dB at the same frequencies. The results therefore confirm stable wideband transmission beamforming across the investigated frequency range.

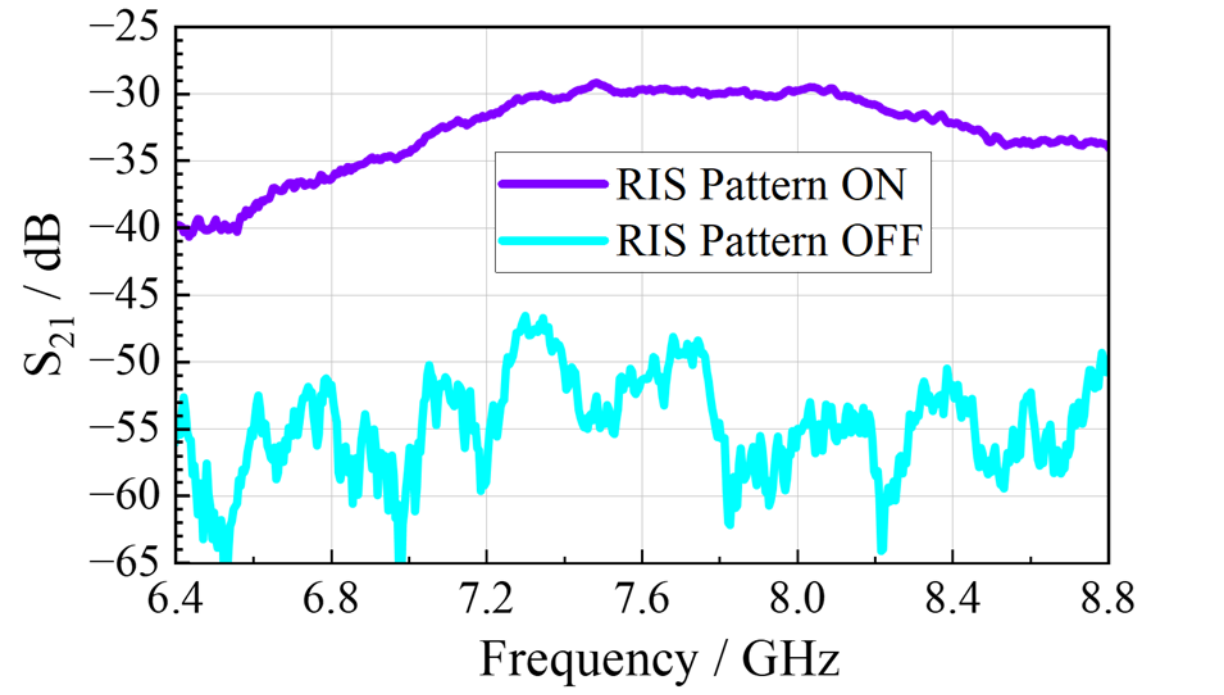

Fig. 21. Measured near-field $S_{21}$ versus frequency under RIS ON and OFF conditions for the transmission mode.

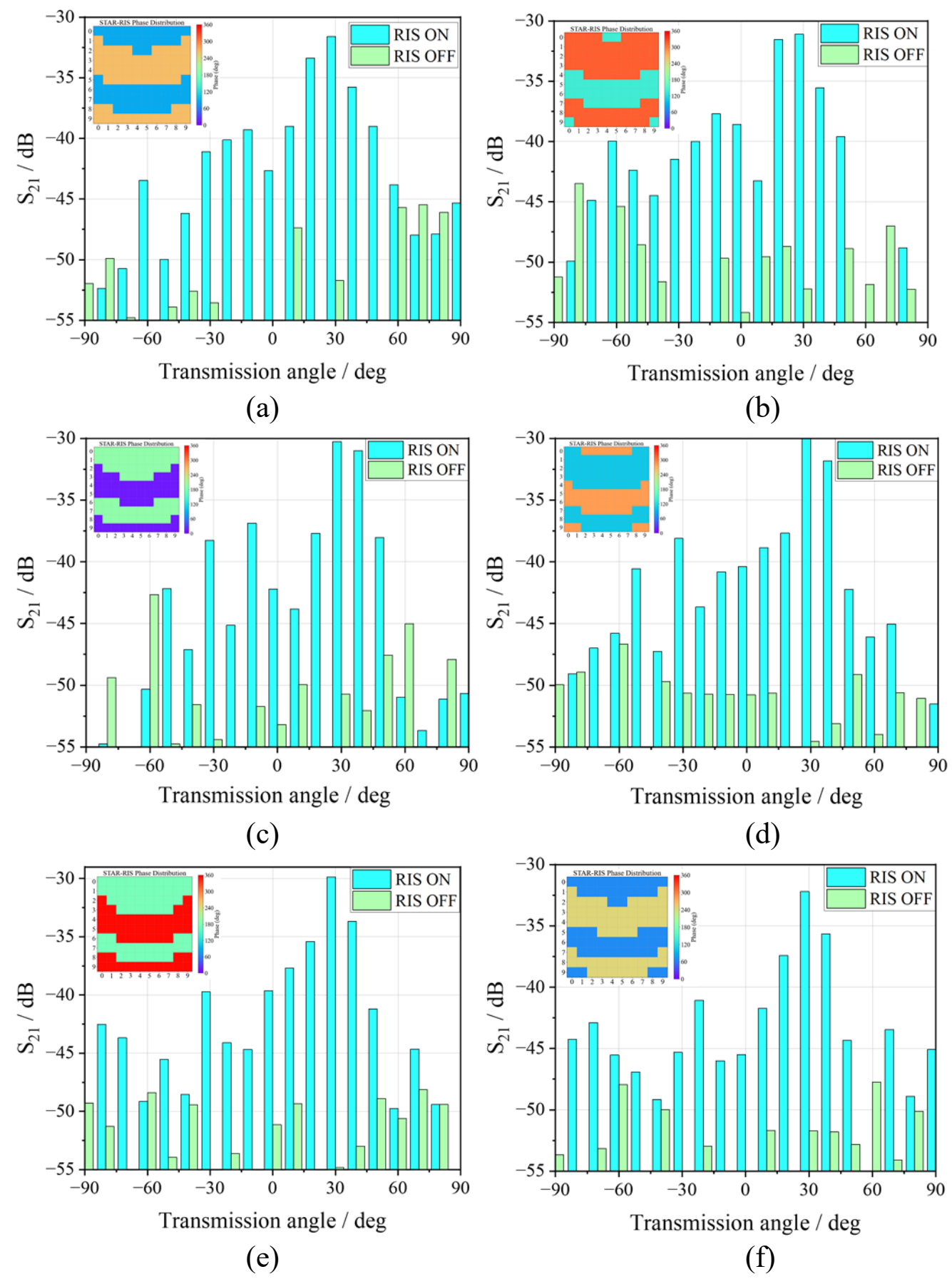

Fig. 22. Measured near-field transmission patterns for a 30° target beam at (a) 6.8 GHz, (b) 7.1 GHz, (c) 7.4 GHz, (d) 7.8 GHz, (e) 8.1 GHz, and (f) 8.4 GHz.

Beam steering on the transmission side is also examined at the center frequency of 7.6 GHz. As shown in Fig. 23, the transmitted beam is reconfigured toward −30°, −15°, 15°, and 30° by updating the applied 1-bit phase distribution while keeping the illumination unchanged. The measured $S_{21}$ values at the four target directions are −28.9, −28.7, −29.6, and −29.7 dB, respectively, with only 1.0 dB variation over the considered scanning range. The corresponding highest sidelobe levels are −37.5, −41.1, −41.0, and −37.2 dB, which are approximately 7.5–12.4 dB lower than the desired beam levels. These measurements demonstrate effective transmission beam steering over an angular range of ±30°. Together with the reflection results in Figs. 16–19, the near-field measurements verify wideband programmable beamforming on both sides of the proposed STAR-RIS.

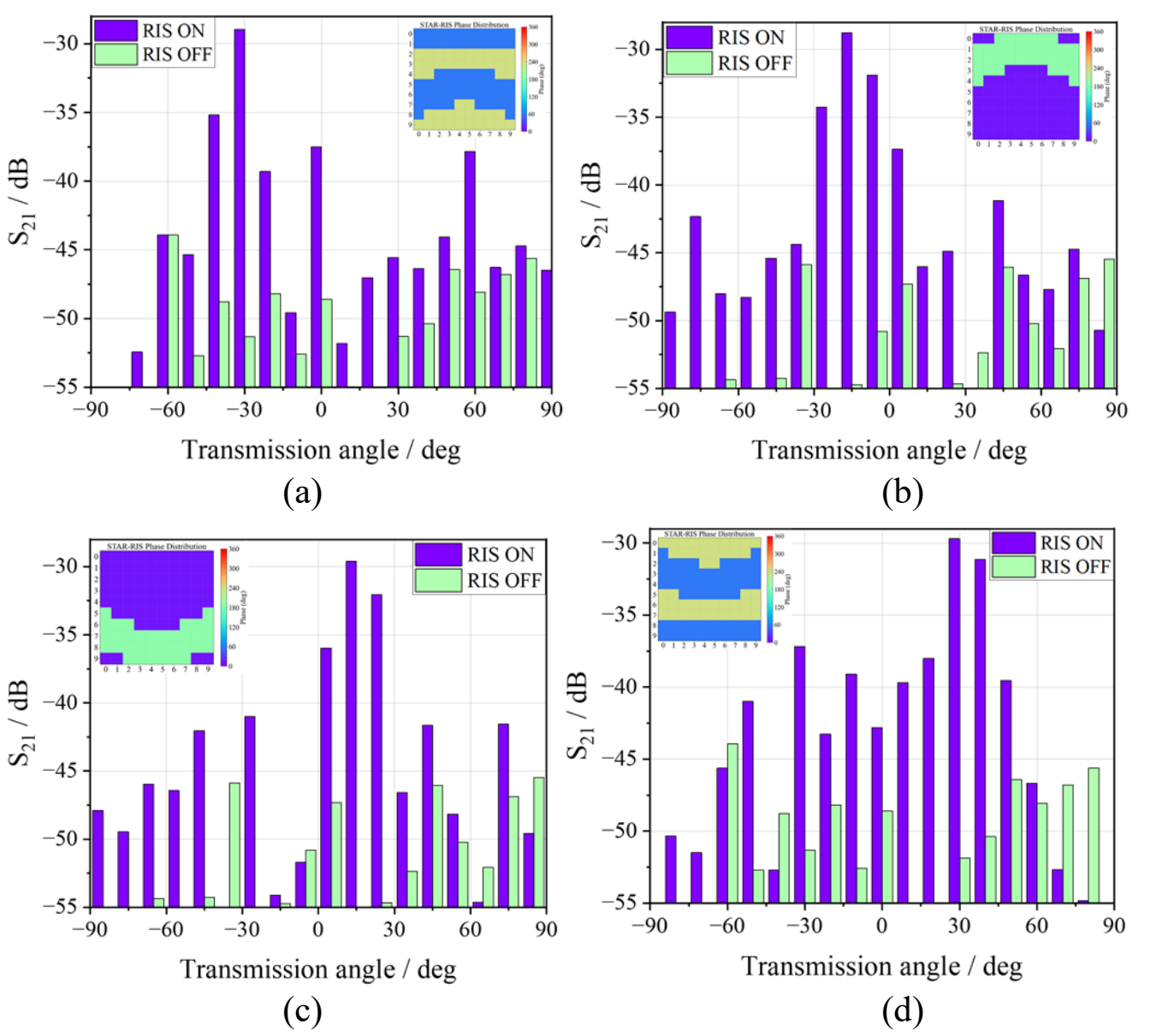

Fig. 23. Measured near-field transmission beam scanning at 7.6 GHz for target angles of (a) −30°, (b) −15°, (c) 15°, and (d) 30°.

## D. *Far-Field Measurement*

To further evaluate the proposed STAR-RIS under far-field propagation conditions, additional measurements are conducted for both reflection and transmission modes. The far-field reflection measurement setup is shown in Fig. 24. An MVG SH-800-277 horn is employed as the transmitting antenna, while an MVG QH-800-33 horn is used as the receiving antenna. The Tx horn is positioned approximately 2 m from the STAR-RIS with normal incidence, and the Rx horn is located approximately 2.3 m from the surface at the desired reflection direction of 30°. The Tx horn is connected to a Keysight VXG M9348B signal generator, whereas the Rx horn is connected to an Agilent EXA N9010A signal analyzer for received-power measurement. These distances place the measurement antennas in the far-field region of the STAR-RIS aperture over the investigated frequency range.

The far-field reflection results at 6.8, 7.1, 7.4, 7.8, 8.1, and 8.4 GHz are presented in Fig. 25. For each frequency, the corresponding 1-bit reflection pattern is applied to steer the reflected beam toward 30°. With the RIS pattern ON, the measured received powers at the desired direction are −61.8, −59.7, −58.1, −60.0, −60.3, and −60.1 dBm, respectively. In comparison, the corresponding RIS OFF levels are −71.7, −71.5, −69.2, −73.7, −73.0, and −67.8 dBm. Therefore, received-power enhancements of 9.9, 11.8, 11.1, 13.7, 12.7, and 7.7 dB are achieved across the six frequency points, with an average enhancement of approximately 11.2 dB. The measured improvement over the entire 6.8–8.4 GHz range further verifies that the wideband reflection behavior is preserved under far-field propagation.

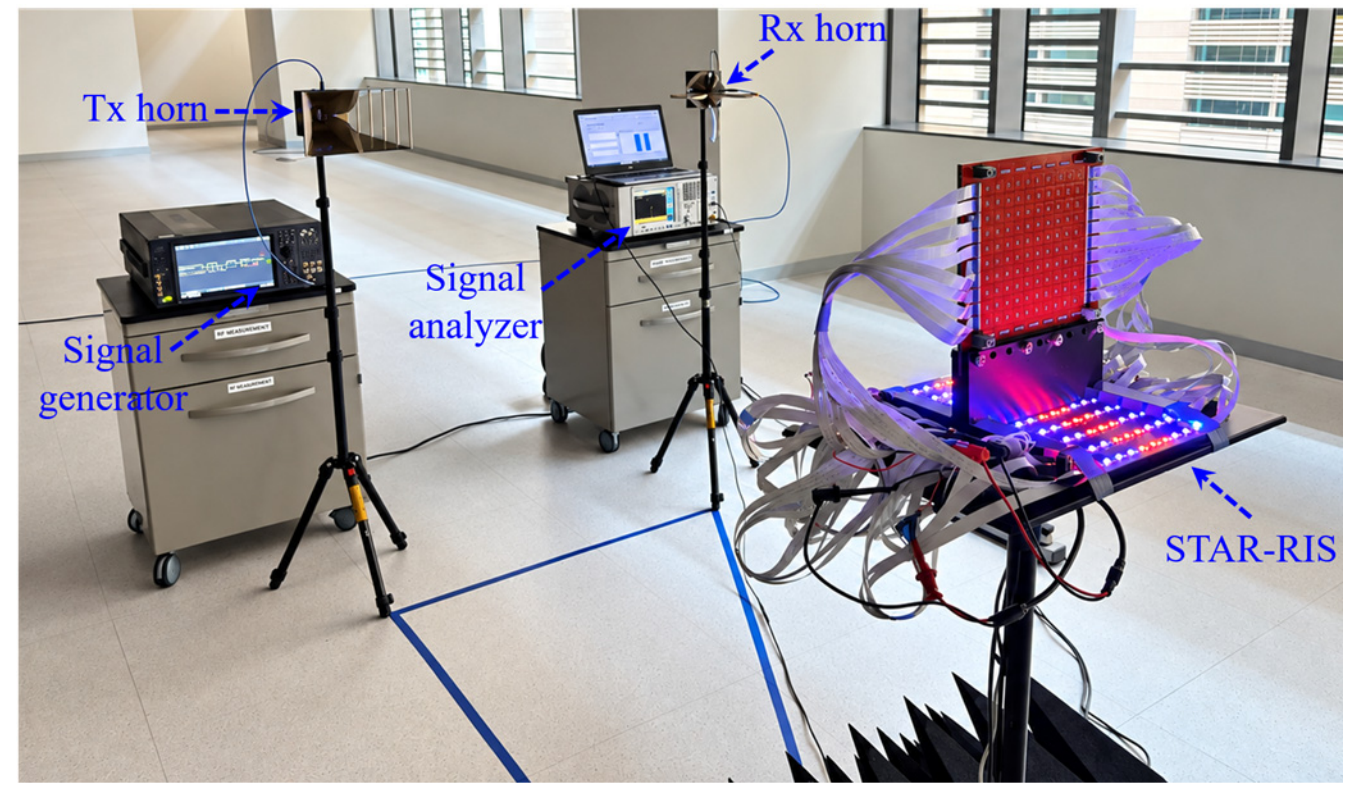

Fig. 24. Far-field measurement setup for the reflection mode.

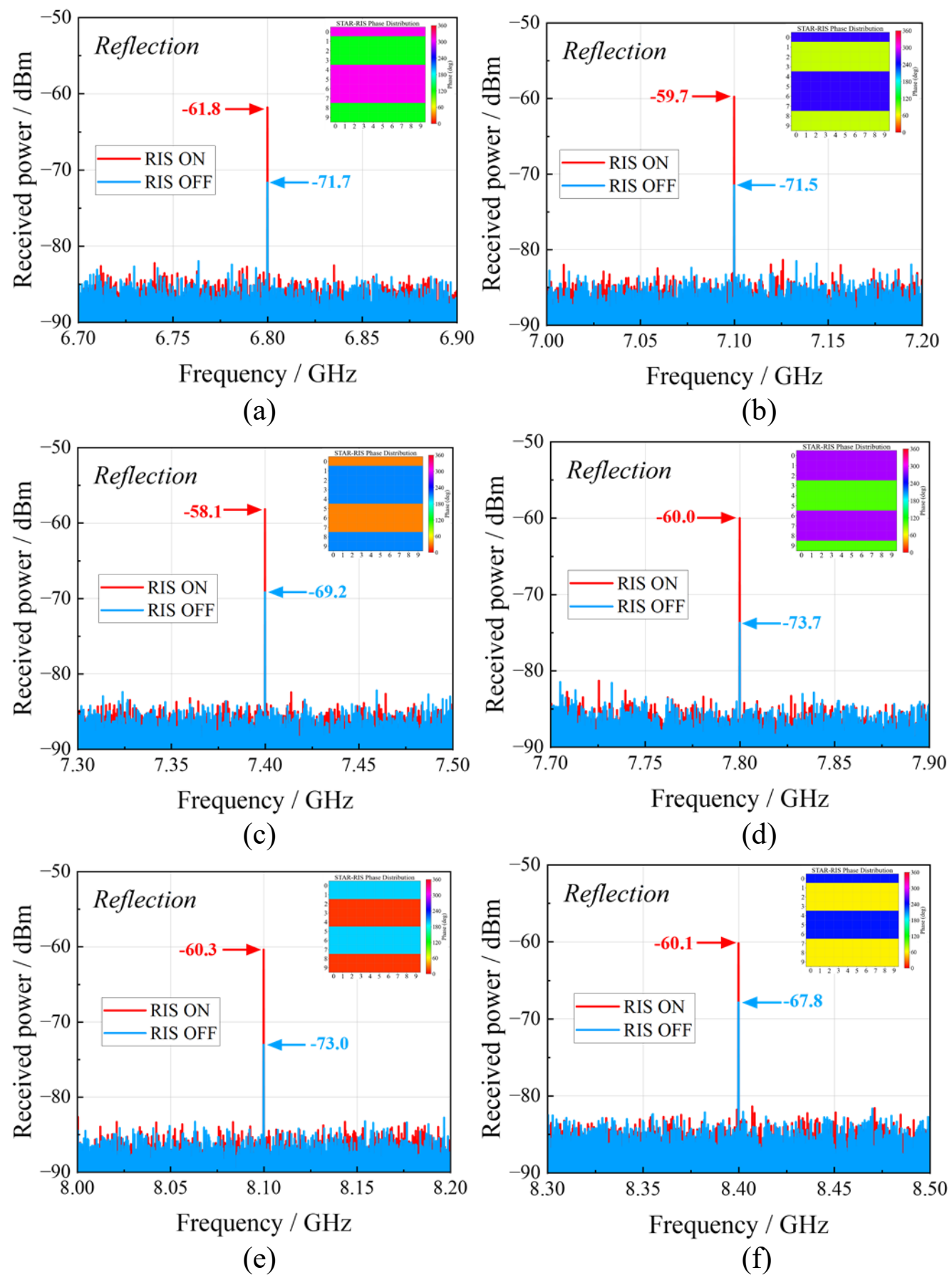

Fig. 25. Measured far-field received power under RIS ON and OFF conditions for 30° reflection at (a) 6.8 GHz, (b) 7.1 GHz, (c) 7.4 GHz, (d) 7.8 GHz, (e) 8.1 GHz, and (f) 8.4 GHz.

For the transmission case, the measurement configuration is modified by moving the Rx horn to the opposite side of the STAR-RIS, as illustrated in Fig. 26. The Tx horn, STAR-RIS, measurement distances, and instrumentation are kept unchanged from the reflection measurement, while the Rx horn is positioned in the opposite hemisphere at the desired 30° transmission direction. This arrangement enables the far-field transmission performance to be evaluated under the same illumination and measurement conditions used for the reflection case.

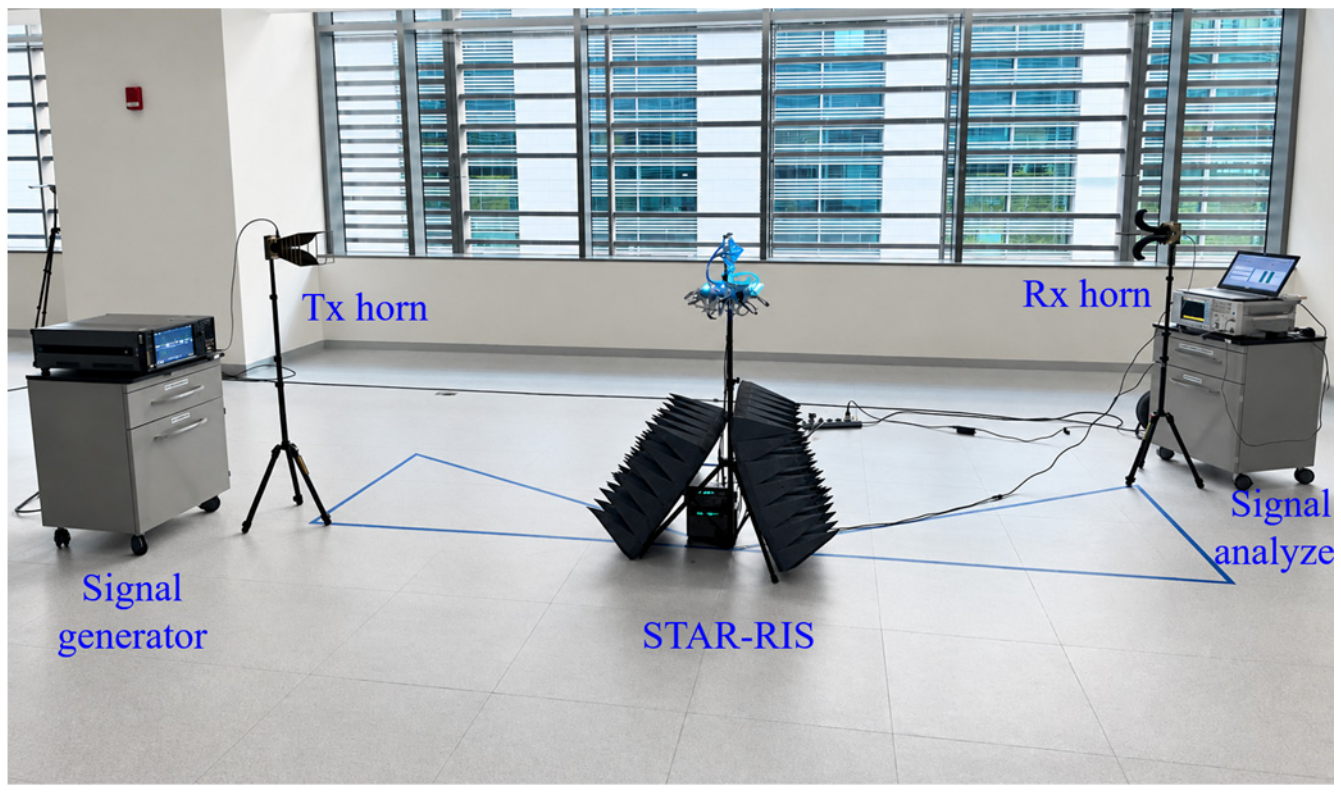

Fig. 26. Far-field measurement setup for the transmission mode.

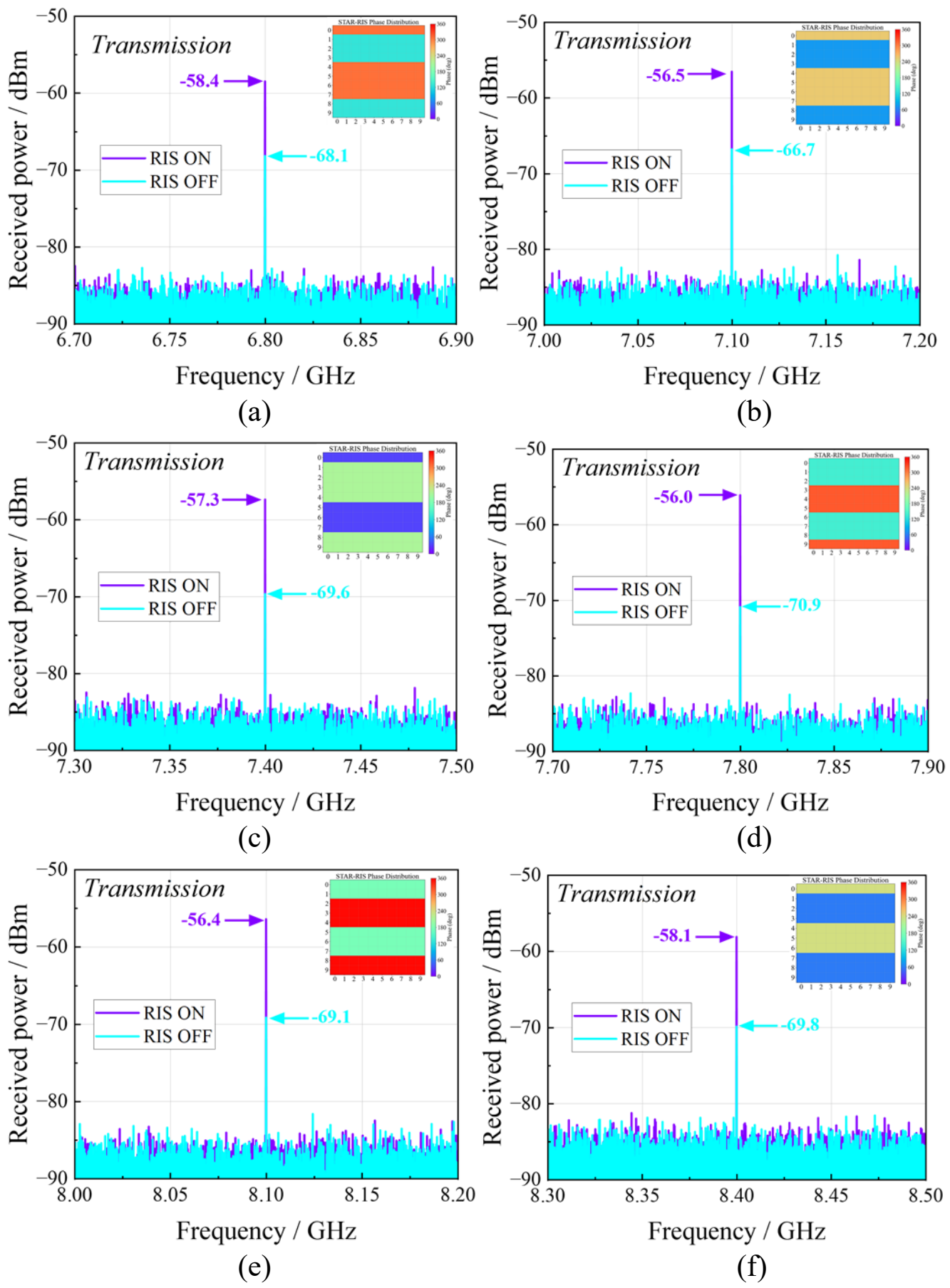

Fig. 27. Measured far-field received power under RIS ON and OFF conditions for 30° transmission at (a) 6.8 GHz, (b) 7.1 GHz, (c) 7.4 GHz, (d) 7.8 GHz, (e) 8.1 GHz, and (f) 8.4 GHz.

Fig. 27 presents the corresponding transmission measurements at the same six frequency points. After applying the frequency-dependent 1-bit transmission patterns, the measured received powers at the desired 30° direction are −58.4, −56.5, −57.3, −56.0, −56.4, and −58.1 dBm at 6.8, 7.1, 7.4, 7.8, 8.1, and 8.4 GHz, respectively. The measured RIS OFF levels are −68.1, −66.7, −69.6, −70.9, −69.1, and −69.8 dBm at the corresponding frequencies. The resulting received-power enhancements are 9.7, 10.2, 12.3, 14.9, 12.7, and 11.7 dB, corresponding to an average enhancement of approximately 11.9 dB. The relatively small variation of the RIS ON received power across the measured band also demonstrates consistent far-field transmission performance over the wide operating frequency range.

Overall, the far-field measurements demonstrate average received-power enhancements of 11.2 and 11.9 dB for reflection and transmission, respectively, across 6.8–8.4 GHz. Together with the near-field results presented in Section IV-C, these measurements verify that the proposed STAR-RIS can realize wideband programmable wavefront manipulation toward both hemispheres under different propagation conditions.

### *E. Comparison and Discussion*

Finally, a comparison with representative practical RIS and STAR-RIS hardware is provided in Table II. Wideband RIS designs have previously been demonstrated in [4] and [7], achieving bandwidths of 26.9% and 13.9%, respectively. However, both designs operate only in the reflection mode, and therefore the programmable coverage is restricted to the same hemisphere as the incident wave. A wideband transmissive RIS was reported in [38], achieving a bandwidth of 14.2%. Although this design redirects the incident wave through transmission, its controllable coverage is still limited to only one hemisphere on the opposite side of the surface. In addition, the transmissive RIS in [38] adopts a 3D configuration with a profile of approximately 1.9$\lambda$, resulting in a bulky structure that is less suitable for low-profile surface-mounted applications such as walls and windows. Therefore, although the RIS designs in [4], [7], and [38] demonstrate wideband operation, they manipulate the EM wave through either reflection or transmission and cannot provide programmable coverage on both sides of the surface.

More recently, practical STAR-RIS hardware capable of manipulating EM waves on both sides of the surface has gradually been developed. The STAR reconfigurable array in [26] achieves reflection and transmission bandwidths of approximately 10% and 8.4%, respectively. However, the DC biasing network is directly exposed on the transmitting side of the surface, which may introduce additional parasitic interactions with the transmitted RF field and also limit the extension toward larger apertures. The active STAR-RIS reported in [27] adopts a planar configuration, but its operating bandwidth is limited to only 1.9% for both reflection and transmission. In [28], a hybrid transmitting and reflecting beyond-diagonal RIS achieves approximately 8.3% bandwidth in both modes. However, its implementation consists of multiple interconnected RF modules and exhibits an overall profile of approximately 0.7$\lambda$, estimated from the reported fabricated prototype, which increases the structural complexity and limits its suitability for low-profile surface integration. Overall, although [26]–[28] demonstrate the feasibility of practical STAR-RIS hardware, their operating bandwidths

TABLE II. PERFORMANCE COMPARISON WITH PREVIOUS REPORTED WORKS

| Ref. | RIS type | RIS Profile | Freq. (GHz) | Numbers of Elements | Array Size (λ) | R/T Bandwidth (%) | Application |
|---|---|---|---|---|---|---|---|
| [4] | Reflective | 2D | 27.5 | 20×20 | 8.0×8.0 | 26.9 | 5G mm-Wave |
| [7] | Reflective | 2D | 26 | 20×20 | 7.1×7.1 | 13.9 | 5G mm-Wave |
| [38] | Transmissive | 3D | 27 | 16×16 | 6.9×6.9×1.9 | 14.2 | 5G and beyond |
| [26] | STAR | 2D | 5.8 | 16×16 | 6.2×6.2 | 10/8.4 | NA |
| [27] | STAR | 2D | 2.6 | 8×4 | 4.0×2.0 | 1.9/1.9 | NA |
| [28] | STAR | 3D | 2.4 | 4×4 | 2.0×2.0×0.7* | 8.3/8.3 | NA |
| This work | STAR | 2D | 7.6 | 10×10 | 3.2×3.2 | 19.8/21.9 | 6G FR3 |

*It donates the observed data based on the authors' reported results.

remain relatively limited, and all of these reported prototypes operate in the sub-6-GHz range.

For future 6G systems, particularly in the emerging FR3 spectrum, a practical STAR-RIS is expected to simultaneously provide wideband operation, a low-profile structure, and a scalable hardware architecture. To address this research gap, this work proposes a planar STAR-RIS operating at 7.6 GHz, achieving measured reflection and transmission bandwidths of 19.8% and 21.9%, respectively. Compared with the existing STAR-RIS hardware summarized in Table II, the proposed design provides substantially wider bandwidth. Moreover, the RF feeding and DC biasing network are embedded between two ground planes, reducing undesired interaction with the external reflection and transmission apertures. Therefore, the proposed design provides a wideband, low-profile, and scalable STAR-RIS hardware platform for programmable reflection and transmission in future 6G FR3 applications.

## V. CONCLUSION

This work presented a wideband planar STAR-RIS. The proposed four-state polarization-converting unit cell realizes 1-bit phase control for both reflection and transmission through electrically controlled current reversal, while the embedded RF/DC network supports a low-profile and scalable implementation. A 10 × 10 prototype and programmable control system were fabricated and experimentally validated, achieving measured 3-dB bandwidths of 19.8% and 21.9% for reflection and transmission, respectively. Near- and far-field measurements further confirm wideband beamforming and programmable control on both sides of the surface. The proposed architecture therefore provides a practical wideband STAR-RIS solution for future 6G FR3 deployment.